\documentclass[sigconf]{acmart}
\usepackage{xcolor}
\definecolor{myColor}{HTML}{E6E6F0} 
\definecolor{myyellow}{RGB}{255, 250, 205} 
\usepackage[noend]{algpseudocode}
\usepackage{multicol}
\usepackage{bbm}
\usepackage{balance}
\usepackage{hyperref}
\usepackage{graphicx}
\usepackage{algorithm}
\usepackage{amsthm}
\usepackage{appendix}
\newtheorem{example}{Example}[section]
\usepackage[english]{babel}
\usepackage{amsthm}
\usepackage{url}
\usepackage{ulem}

\usepackage{mdframed}
\usepackage{algorithmicx}
\usepackage{subfigure}
\usepackage{multirow} 
\AtBeginDocument{%
  }
\newmdenv[
  backgroundcolor=myyellow,
  roundcorner=5pt,  
  linewidth=0pt     
]{yellowbox}

\setcopyright{acmlicensed}
\copyrightyear{2018}
\acmYear{2018}
\acmDOI{XXXXXXX.XXXXXXX}
\acmConference[Conference acronym 'XX]{Make sure to enter the correct
  conference title from your rights confirmation email}{June 03--05,
  2018}{Woodstock, NY}
\acmISBN{978-1-4503-XXXX-X/2018/06}

\begin{document}

\title{Algorithmic Complexity Attacks on All Learned Cardinality Estimators: A Data-centric Approach}
\author{Yingze Li, Xianglong Liu, Dong Wang, Zixuan Wang, Hongzhi Wang$^*$, Kaixin Zhang, Yiming Guan}
\affiliation{%
  \institution{Harbin Institute of Technology}
  \country{China}
}
\email{{23B903046}@stu.hit.edu.cn, wangzh@hit.edu.cn}

\begin{abstract}
Learned cardinality estimators show promise in query cardinality prediction, yet they universally exhibit fragility to training data drifts, posing risks for real-world deployment. This work is the first to theoretical investigate how minimal data-level drifts can maximally  degrade the accuracy of learned estimators. We propose data-centric algorithmic complexity attack against learned estimators in a black-box setting, proving that finding the optimal data drift that compromise all estimators is NP-Hard. To address this, we design a polynomial-time approximation algorithm with a $(1-\kappa)$ approximation ratio.  
Extensive experiments demonstrate our attack's effectiveness: on STATS-CEB and IMDB-JOB benchmarks, modifying just 0.8\% of training tuples increases the 90-th percentile Qerror by three orders of magnitude and raises end-to-end processing time by up to 20$\times$. Our work not only reveals critical vulnerabilities in deployed learned estimators but also provides the first unified worst-case theoretical analysis of their fragility under data drifts. Additionally, we identify two countermeasures to mitigate such black-box attacks, offering insights for developing robust learned database optimizers.\looseness=-1


\looseness=-1 
\end{abstract}




\maketitle

\section{Introduction}


\uline{C}ardinality \uline{E}stimation (CE) plays a crucial role in query optimization within database systems, as it predicts the number of tuples that satisfy a query without execution and aids the database optimizer select the best query execution plan with the lowest algorithmic time complexity~\cite{SagaDB}, thus improving system efficiency\cite{AreWeReady4CE,DBMSCE}. Compared to traditional estimators~\cite{larson2007SampleCE1,ViswanathPoosala1996ImprovedHF}, learned cardinality estimators offer more accurate estimations and have therefore garnered extensive attention in recent years~\cite{yang2019deep,ALECE,hilprecht2019deepdb,MSCN_Kipf2018LearnedCE}. These learned approaches are primarily categorized into three types: (1) Data-driven approaches~\cite{yang2019deep,hilprecht2019deepdb}, which learn joint data distributions to enable accurate and generalized estimations that are robust to changes among query distributions. (2) Query-driven approaches~\cite{MSCN_Kipf2018LearnedCE,LWNN}, which employ regression models to directly map query representations to cardinalities. (3) Hybrid approaches~\cite{ALECE,wu2021unified,RobustMSCN}, which combine the information from both data and queries to provide a comprehensive prediction. 




Despite the proven superior performance of learned estimators and their potential to replace traditional methods, the robustness and security concerns with these learned approaches have garnered widespread attention~\cite{PACE,RobustMSCN,ALECE,DBMSCE}. In fact, similar concerns arise not only in the CE domain but also across numerous learned database topics~\cite{PoisonLI,ACA_Learnedindex,ATTACK_IndexAdvisor}. When the training data are poisoned~\cite{PoisonLI,PACE,ACA_Learnedindex,ATTACK_IndexAdvisor} or the testing data are inconsistent with the training data~\cite{RobustMSCN,ALECE}, the performance of these learned models can degrade catastrophically. Therefore, studying and understanding how to bring the most extreme performance degradation to these learned DB components with adversary input, known as \uline{A}lgorithm \uline{C}omplexity \uline{A}ttacks (ACA) on these learned database components, can help us develop stronger preventive measures within learned databases and avoid the worst-case scenario~\cite{ACA_Learnedindex,ATTACK_IndexAdvisor}.\looseness=-1




Query-centric ACA has been devised to test the robustness of query-driven CE models.  PACE~\cite{PACE}, proposed by Zhang et al., involves adding noise to query workloads and poisoning the query-driven estimators. This degrades the accuracy of query-driven models and slows down the end-to-end performance. Although query-centric attacks can significantly impair the performance of query-driven estimators, we find that they have limited impact on widely used data-driven learned  estimators~\cite{yang2019deep,nec,hilprecht2019deepdb,CardIndex,FactorJoin,Flat,BayesCard,ASM,wang2021face}. This is primarily because data-driven methods utilize data-level information that is independent of historical workloads to achieve robust cardinality estimation across diverse testing query distributions. Consequently, these methods can effectively withstand poisoning attacks that target historical workloads. Therefore, query-centric ACA methodology is not suitable for test the robustness of these prevalent data-driven models.\looseness=-1







\begin{figure}[htbp]
	\centering
	\includegraphics[width=7.5cm]{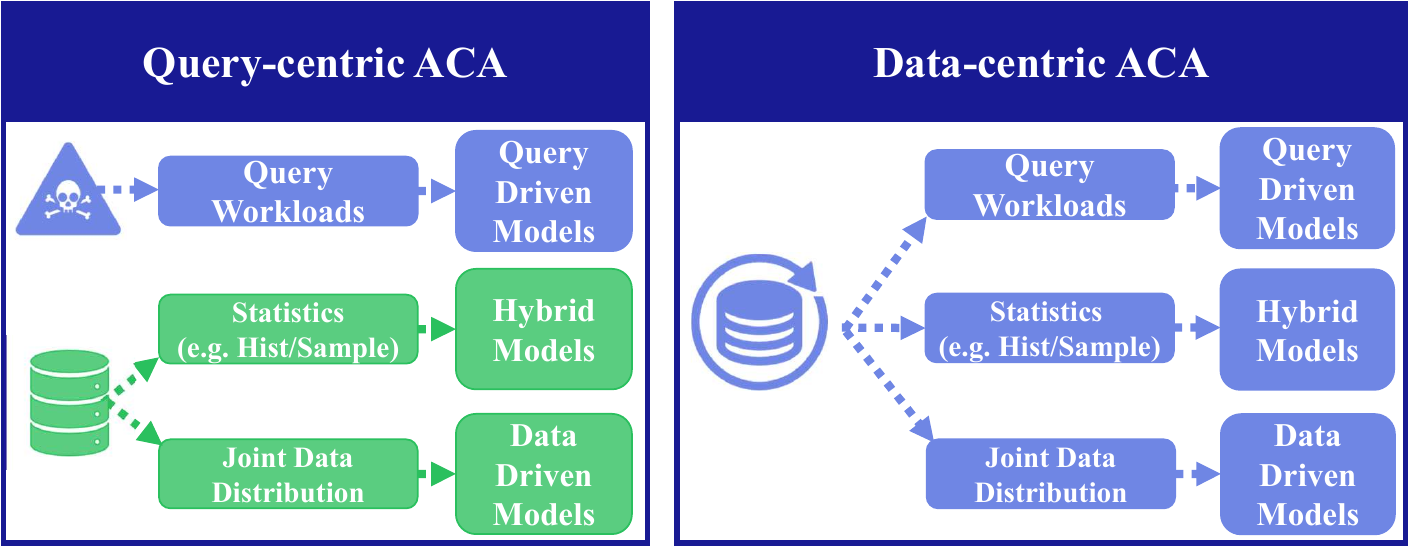}
	\caption{ Query-centric ACA (Left) vs. data-centric ACA (Right): green indicates clean, blue indicates attacked. \looseness=-1}
	\label{P1}
 \vspace{-1em}
\end{figure}





%
Additionally, changes in database can influence the performance of nearly all cardinality estimators~\cite{LuyaoSIGMOD22,wang2021face,ALECE,unlearn,DDU_SIGMOD23}. Fundamentally, all existing learned estimators are learning from statistical information based on their training data sources~\cite{CardIndep}, such as cardinality labels from historical workloads~\cite{LWNN,MSCN_Kipf2018LearnedCE}, or joint data distributions within the training dataset~\cite{nec,hilprecht2019deepdb,FactorJoin}. When the training information becomes outdated, the underlying statistical properties inherently shift, causing all models trained on these stale statistics to incorporate erroneous information~\cite{ALECE,DDU_SIGMOD23}. In other words, in dynamic database environments, any data update effectively constitutes an "attack" on outdated learned estimators, leading to performance degradation. Given the ubiquity of dynamic database scenarios in practice, this widespread attack surface motivates us to role-play an adversary attacker to design  a \textbf{data-centric black-box ACA} that simulates worst-case situation where \textbf{minimum data drifts can maximally compromise the performance of all outdated learned estimators.} \looseness=-1


However, initiating an optimal black-box ACA via a data-centric approach involves overcoming the following key challenges: (1) \textbf{Black-box Limitations.} The attacker lacks information about the currently deployed cardinality estimator and does not know whether it is MLP~\cite{LWNN,yang2019deep}, CNN~\cite{MSCN_Kipf2018LearnedCE,RobustMSCN}, Transformer~\cite{ALECE,nec}, or Bayes Network~\cite{BayesCard,FactorJoin}. This uncertainty restricts the attacker's options and prohibits the attacker from exploiting the existing white-box attack strategies designed for compromising specific machine learning models~\cite{PoisonLI,ACA_Learnedindex,ATTACK_IndexAdvisor,PACE}. (2) \textbf{Tractability Awareness.} 
\textcolor{black}{ Can the maximum damage of this attack be pre-calculated in polynomial-time? If computationally feasible, such analysis allows us to proactively evaluate the worst-case behavior of all learned estimators, acting as a theoretical `canary in a coal mine' to alert us to potential worst-case scenarios in advance.} (3) \textbf{Limited Budget.} How can a limited number of attack operations on the training data maximally degrade the accuracy of models trained on this poisoned dataset? On the one hand, excessive data modification operations  may take long to perform, and during this period, a new updated model could potentially be trained based on the new distribution. On the other hand, small-scale data modifications seem to have limited impact on learned estimators. For instance, prior studies have shown that outdated models can tolerate data-level changes exceeding 30\% while still maintaining strong estimation performance~\cite{AreWeReady4CE,unlearn,hilprecht2019deepdb}.\looseness=-1    



According to the above discussions, the toughest challenge lies in that the attacker has no information about the estimator deployed internally within the black-box, which hinders the attacker from effectively quantifying and boosting the efficiency of the attack. To address this issue, we discovered a crucial and often overlooked fact: learned estimators are approaching oracle-level accuracy on training datasets, obtaining accurate cardinality. For example, many learned estimators~\cite{Flat,yang2019deep,ALECE,BayesCard,wang2021face,hilprecht2019deepdb} achieve more than 90\% of the Qerrors close to 1. However, this primary advantage also exposes these models to a common weakness. Despite their varying model types, the predictive output behavior of each model tends to closely mirror the oracle's behavior within the training dataset.\looseness=-1




Therefore, we can leverage this exposed common weakness to address the aforementioned challenge. Instead of directly tackling the black-box attack head-on and lacking the necessary information to solve the problem, we can indirectly construct a finite set of data drifts designed to poison the oracle in the training database, and thereby impact the performance of all learned estimators. Thus, to overcome the first black-box challenge, we utilize the oracle based on the training data as our surrogate model and convert the difficult black-box attack problem into a tuple selection problem with much more information.  To tackle the second tractability challenge, we formulate this data-centric attack as a combinatorial optimization problem and establish that this problem is NP-Hard.  To achieve the most effective attack within a limited budget and address the third challenge, we analyze the supermodular properties of the optimization objective under specific conditions and design a $(1-\kappa)$ approximation algorithm for efficient resolution.

\textbf{Contributions:} We make the following contributions.

C1: \textbf{Black-box Attack on All Learned Estimators}: We investigate how to launch a successful \uline{D}ata-centric  \uline{A}lgorithm \uline{C}omplexity \uline{A}ttack (DACA) that uses minimum data drifts to maximally compromise the performance of all outdated learned estimators in a black-box setting (\S~\ref{ATK_ORC}).

C2: \textbf{Intractability Insights}: We demonstrate that, under a limited budget, finding the optimal  data drift that compromise all estimators is NP-Hard, even when only considering deletion adversary operations (\S~\ref{Sec.CC}).

C3: \textcolor{black}{\textbf{Near-Optimal Attack Strategy}: We design a polynomial-time approximation algorithm (\S~\ref{sec.AUAF}) that guarantees a $(1 - \kappa)$ approximation for the worst-case performance of all estimators. This analysis acts as a theoretical  early warning mechanism, allowing us to proactively quantify and approximate the worst-case scenarios of all learned estimators (\S~\ref{sec.AAS}). }




C4: \textbf{Common Security Vulnerabilities}:  Our experiments (\S~\ref{sec.exp}) show that perturbing a single database's data distribution by merely 0.8\% significantly degrades learned cardinality estimators trained on poisoned data, causing:  a 1000$\times$ increase in 90-th percentile Qerror, and  up to $20\times$ longer end-to-end processing time. This exposes a fundamental vulnerability in most learned estimators. We further propose practical countermeasures to defense such black-box attacks, advancing robust learned database optimizers.\looseness=-1

\vspace{-1em}

\section{PRELIMINARIES}
\label{Sec.PRELIMINARIES}

We present   learned cardinality estimators in \S~\ref{sec.lce},  algorithmic complexity attacks against learned DBMS components in \S~\ref{Sec.ACA}, and our formal threat model in \S~\ref{sec.threatModel}.


\vspace{-1em}

\subsection{Learned Cardinality Estimation}
\label{sec.lce}

Suppose a database $D$ consists $\ell$ relations, i.e. $D = \{ R_1,R_2,\dots R_{\ell}\}$. Given a query $\textbf{q}$ ,  when $\textbf{q}$ is executed on $D$, we obtain \textbf{q}'s results in result set $\mathbbm{S}(\textbf{q}|D)$.  We  denote the cardinality of $\mathbbm{S}(\textbf{q}|D)$ as $ C_D(\textbf{q})$.\looseness=-1 


\textbf{Tuple's Joint Weight}: For relation $R_x$ with $N$ tuples, a tuple $t_i\in R_x$ and a query $\textbf{q}_j$, the joint weight $w_{ij}$ denotes tuple $t_i$'s contribution to query $\textbf{q}_j$'s result's cardinality, that is $w_{ij} =  |\mathbbm{S}(\textbf{q}_j|D-R_{x}+\{t_i\})|$. $\textbf{q}_j$'s cardinality can be seen as a linear summation of $R_x$'s tuple's the joint weights: $C_D(\textbf{q}_j) = \sum_{i=1}^N w_{ij}$.

\textbf{Learned Cardinality Estimation}: Learned cardinality estimation needs to make predictions on $C_D(\textbf{q})$ using  learned estimator $E$ without executing $\textbf{q}$ on $D$. $E$ is trained via the information of $D$. 
Based on the training information, these learned cardinality estimators can be categorized into these main paradigms: data-driven, query-driven, and hybrid.




\textbf{Learned Data-driven Estimator}: Learned data-driven cardinality estimator $E_{Data}$ learns the joint distribution of database $D$. Based on the learned statistical information, $E_{Data}$ provides the query independent estimation of the query \textbf{q}'s cardinality $est(\textbf{q}|E_{Data})$.

\textbf{Learned Query-driven Estimator}: Learned  query-driven cardinality estimator $E_{Query}$ learns the mapping from historical workloads $\mathcal{Q}$ to  cardinalities $\mathcal{C}$, i.e., $\mathcal{Q}\rightarrow\mathcal{C}$. Based on the learned mapping knowledge, $E_{Query}$  provides the estimation of the query \textbf{q}'s cardinality $est(\textbf{q}|E_{Query})$.

\textbf{Hybrid  estimator}: Learned hybrid estimator $E_{Hybrid}$ combines query $\mathcal{Q}$ with  statistics (e.g. histograms~\cite{ALECE}, samples~\cite{MSCN_Kipf2018LearnedCE}) $\mathcal{S}$ as the input features, and learns the mapping $\{\mathcal{Q} \times \mathcal{S}\} \rightarrow\mathcal{C}$. Based on the enhanced features, $E_{Hybrid}$ provides a comprehensive estimation of  query \textbf{q}'s cardinality $est(\textbf{q}|E_{Hybrid})$ compared to $est(\textbf{q}|E_{Query})$.\looseness=-1


For query $\textbf{q}_j$, we follow the assumptions made in the majority of CE literature~\cite{ALECE,nec,ASM,DBMSCE,MSCN_Kipf2018LearnedCE}, considering $\textbf{q}_j$ to be the most prevalent selection-projection-join query of the form:
\begin{center}
\texttt{SELECT * FROM } $\textbf{q}_j.{Tabs}$ \texttt{ WHERE } $\textbf{q}_j.{Joins}$ \texttt{ AND } $\textbf{q}_j.{Filters}$
\end{center}

where: $\textbf{q}_j.{Tabs}$ denotes the set of tables involved in the query $\textbf{q}_j$, such that $\textbf{q}_j.{Tabs} \subseteq {D}$, and  $\textbf{q}_j.{Joins}$ represents the join predicates,  $\textbf{q}_j.{Filters}$ signifies the filter predicates.

\begin{figure*}[htbp]
	\centering
	\includegraphics[width=13 cm]{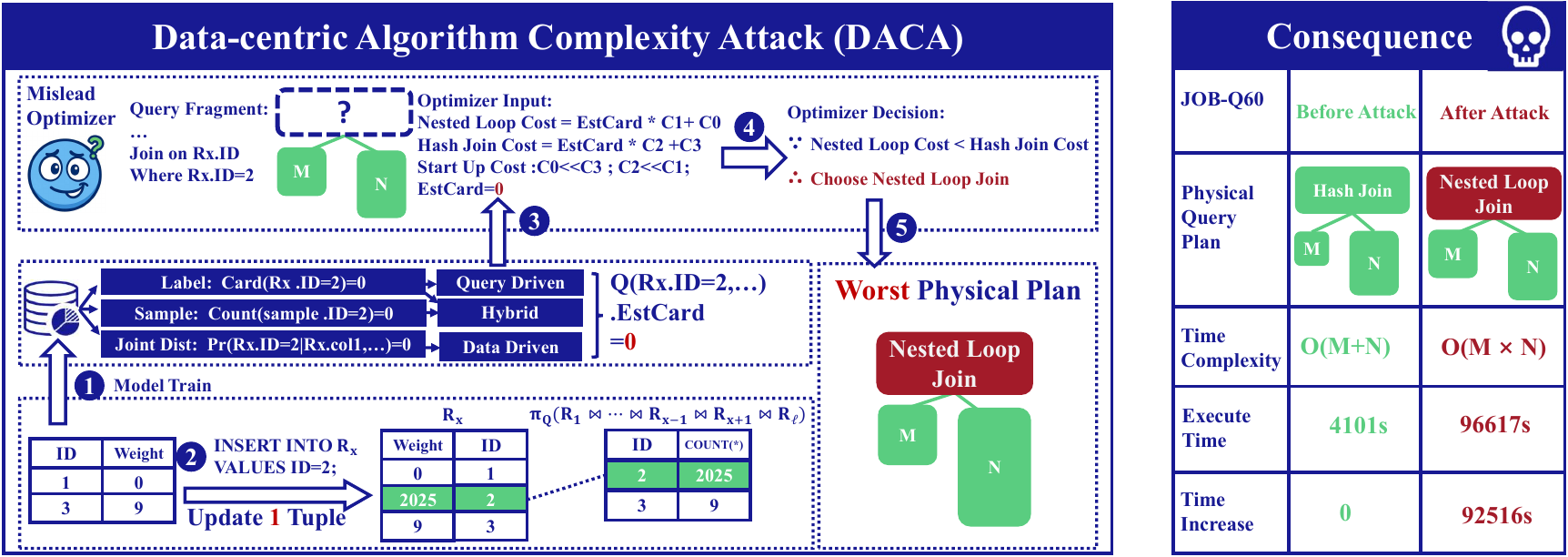}
         \vspace{-1em}
	\caption{DACA workflow (Left); The consequence of DACA (Right).}
	\label{PMerge}
 \vspace{-1em}
\end{figure*}
\subsection{Algorithmic Complexity Attacks}
\label{Sec.ACA}


The \uline{A}lgorithmic \uline{C}omplexity \uline{A}ttack (ACA) is a variant of Denial-of-Service attacks~\cite{DDOSACA,ACA_SIGCOMM,ACA_SIGSAC,DDOS}. In this attack, the adversary introduces minimal adversarial perturbations into the target system, altering its computational complexity and exhausting computational resources. ACA has been extensively employed to assess the worst-case robustness of learned database components~\cite{ACA_Learnedindex,PACE,PoisonLI}. For example, in the ACA against dynamic learned indexes~\cite{ACA_Learnedindex}, Kornaropoulos et al. designed an adversarial index insertion strategy that increased memory overhead and query time complexity, triggering an out-of-memory error with only a few hundred adversarial insertions.\looseness=-1

Fundamentally, the database optimizer \textit{\textbf{decreases}} the computational complexity of physical plans by utilizing cost estimates provided by estimators and applying relevant rewrite rules~\cite{SagaDB}. However, if the estimator is compromised by misleading training information, it may mislead the optimizer to select the worst execution plan, thereby \textit{\textbf{increasing}} the time complexity of the physical plan. Targeting this vulnerability, PACE~\cite{PACE} achieves a query-centric ACA by contaminating the historical query workload, thereby affecting query-driven estimators. However, PACE is unable to influence data-driven estimators, as these models rely on data features that are independent of the historical workload for their predictions, rendering them unaffected by the poisoned historical workload.
Surprisingly, we find that in dynamic database environments, even minor data drifts can constitute a powerful \textit{data-centric ACA} on stale estimators. This attack simultaneously impacts \textit{all} outdated estimators, regardless of their underlying architecture, and  greatly increases the computational complexity of physical plans, as demonstrated in the example below:\looseness=-1
%


\begin{example}
Figure~\ref{PMerge} shows one possible data-centric ACA on learned estimators by \textbf{updating a single tuple} from the original database. We assume that the  estimators have already been trained on the previous database (Step~1). In our scenario, only a single tuple with $R_x.ID=2$ is inserted (Step~2).  While a single tuple update does not trigger the re-finetuning condition of the cardinality estimator, such a small but overlooked update is already sufficient to affect the knowledge of all estimators, whether the cardinality labels in historical queries, sampling points within a sampling pool, or joint data distributions in databases,   will all be aligned with a cardinality label of 0. Thus, almost all estimators fall into  predicting a cardinality of 0 for queries containing predicates on $R_x.ID=2$ (Step~3).   Incorrect cardinality estimates distort the estimated costs, causing the optimizer to erroneously assume that the cost of a Nested Loop join is lower than that of a Hash Join (Step~4). Consequently, the optimizer opts for the Nested Loop operator (Step~5), which severely amplifies the query’s theoretical time complexity and physical execution time. Figure~\ref{PMerge} (right) shows one possible consequence on JOB-Q60, the  complexity of the query operator degrades from Hash Join’s \(O(M + N)\) to Nested Loop’s \(O(M \times N)\), and the execution time increases $9.2\times 10^{4} $ seconds.\looseness=-1  
\end{example}

\vspace{-1em}

\subsection{Threat Model}
\label{sec.threatModel}

In this section, we will establish the threat model via our adversary analysis. We formalize the threat model by defining the adversary, their goals, knowledge, and attack evaluation metrics.\looseness=-1


\textbf{Adversary and Adversary's Goal:} To investigate the worst-case performance of all cardinality estimators in dynamically changing databases, we postulate a virtual adversary that emerges when the estimator's information becomes outdated. To uniformly analyze the impact of different attack strategies under consistent distributions, we fix the updated database state as $D$. The adversary aims to find minimized adversarial modifications $\Delta D$ (insertions/deletions) such that models trained on the stale distribution $D' = D + \Delta D$ yield catastrophically inaccurate estimations  when applied to the latest version $D$. This misleads the optimizer to select physical plans with the highest time complexity when evaluated on the latest version $D$. For brevity, we refer to $D$ as the \textit{cleaned state} and $D'$ as the \textit{poisoned state}.\looseness=-1



\textbf{Adversary’s Knowledge:} In this work, we study the  black-box attack on learned cardinality estimators. The attacker has no information about the estimator used in the database, except that it is a learned  estimator trained on database $D'$. The attacker is unaware of the specific parameter details of the model and does not know whether it is data-driven or query-driven. Meanwhile, apart from the testing workloads $W_{Test}$ and their results $\mathcal{R}es = \{\mathbbm{S}(\textbf{q}_j|D), \textbf{q}_j\in W_{Test}\}$, the attacker does not know any other data distribution within this database.\looseness=-1


\textbf{Adversary’s Capacity:} We assume that the attacker can access the testing workloads $W_{Test}$ and their results $\mathcal{R}es$. Furthermore, the attacker is capable of performing projection and group-by operations on $\mathcal{R}{es}$ to obtain the joint weights $w_{ij}$ of each tuple $t_i$ in table $R_{x}$ that participates in the testing query $\mathbf{q}_j$. Additionally, 
we assume that the attacker can execute at most $K$ update operations $\Delta D$ (insertions/deletions) on a specific relation $R_x$ within the training database, poisoning the training data into $D' =D+\Delta D$. \looseness=-1




\textbf{Attack Evaluation Metric.}  The attacker has to devise specific strategies to impair these commonly used  metrics: (1) \textbf{Qerror metric}: Defined as $Q(est(\textbf{q}|E), C_D(\textbf{q}))=max(\frac{est(\textbf{q}|E)}{C_D(\textbf{q})},\frac{C_D(\textbf{q})}{est(\textbf{q}|E)})$, where $est(\textbf{q}|E)$ is the estimators $E$'s prediction on given query \textbf{q}. It measures the distance between the estimated cardinality $est(\textbf{q}|E)$ and the true cardinality $C_D(\textbf{q})$ of a query. (2) \textbf{End-to-end latency}: Measures the total latency  for generating a plan with estimated cardinalities and executing the physical plan. \textcolor{black}{This metric is essential for demonstrating how an estimator can improve query optimization performance. It serves as a gold standard for assessing the effectiveness of CE approaches~\cite{DBMSCE,FactorJoin,ALECE,ASM}}. 


\section{Black-Box Attacks against All Learned Estimators}

\label{Sec.BB}


In this section, we will apply the  concepts of algorithmic complexity attack and threat model established in \S~\ref{Sec.PRELIMINARIES} to define the black-box data-centric algorithmic complexity attacks in \S~\ref{sec.PD}. We will appropriately transform this problem into a constrained integer nonlinear programming problem in \S~\ref{ATK_ORC} and analyze its computational intractability in \S~\ref{Sec.CC}.\looseness=-1




\subsection{Black-box Data-centric ACA}
\label{sec.PD}




In this section, we  provide the original definition of black-box data-centric ACA and introduce the challenge associated with directly solving this black-box problem.  We assume that the attacker can execute at most $K$ tuple-level data modifications $\Delta D$ on the clean database state $D$. \textcolor{black}{Thereby poisoning the training state into $D'= D + \Delta D$}. The attacker aims to ensure that, when evaluated on the testing workload $W_{Test}=\{ \textbf{q}_j| 1\leq j \leq M \} $ with $M$ queries, the estimator $E$, trained on the poisoned database $D'$, has the worst accuracy on the original database $D$, i.e., to maximize Eq.~\ref{Eq001}: \looseness=-1


\begin{align}
    \label{Eq001}
    & \text{Max}: \sum_{j=1}^{M}Q(est(\textbf{q}_j|E), C_D(\textbf{q}_j)) =  \sum_{j=1}^{M}\max\left( \frac{est(\mathbf{q}_j|E)}{C_D(\textbf{q}_j)}, \frac{C_D(\textbf{q}_j)}{est(\mathbf{q}_j|E)} \right) \\
    & \text{s.t. } (1) D' = D + \Delta D \quad (2) E\text{ is trained on } D'\ \quad (3)   |\Delta D| \leq K\notag\\ \notag 
\end{align}

When maximizing Eq.~\ref{Eq001}, we assume a complete black-box attack scenario, where the attacker lacks knowledge of the internal estimator type used in the database. Specifically, while the attacker knows that $E$ is trained on $D'$, they are unaware of whether $E$ is data-driven, query-driven, or hybrid. Additionally, the black-box setting restricts the attacker's visibility by concealing details of the deployed model, introducing significant challenges. The internal models could include MLPs~\cite{LWNN,yang2019deep}, CNNs~\cite{MSCN_Kipf2018LearnedCE,RobustMSCN}, Transformers~\cite{ALECE,nec}, or Bayesian networks~\cite{BayesCard,FactorJoin}. Consequently, the attacker cannot leverage existing white-box attack methodologies from ML security~\cite{ACA8,ACA29,ACA65,PACER0,PACER1,PACERX,PACERY}. This renders direct black-box data-centric ACA seemingly infeasible, necessitating a novel approach to reframe the problem.\looseness=-1 





\subsection{Attack the Surrogate Oracle}
\label{ATK_ORC}



In this section, we will take a different perspective to transform the seemingly impossible black-box ACA problem from the previous section into a constrained integer nonlinear programming problem that can be tackled. Specifically, we are inspired by the experimental results of learned CE studies~\cite{yang2019deep,nec,ALECE,AreWeReady4CE,DBMSCE}, which show that existing learned estimators possess extremely high estimation accuracy, with their estimates on the training database approaching oracle-level. Therefore, we utilize an oracle estimator $E_{O}$ on the dataset \( D' \) as our surrogate model. We assume that the $E_{O}$ trained on \( D' \) can precisely report the actual cardinality of queries in \( D' \). For $E_{O}$, we have \(  {est}(\mathbf{q}|E_{O}) = C_{D'}(\mathbf{q})  \). Thus, our task now is to maximize the Qerror of the oracle trained on $D'$:\looseness=-1


{\small
\begin{align}
& \text{Max}:  \sum_{j=1}^{M}   \max\left( \frac{{est}(\mathbf{q}_j|E_{O})}{C_D(\mathbf{q}_j)}, \frac{C_D(\mathbf{q}_j)}{{est}(\mathbf{q}_j|E_{O})} \right) = \sum_{j=1}^{M}   \max\left( \frac{C_{D'}(\mathbf{q}_j)}{C_D(\mathbf{q}_j)}, \frac{C_D(\mathbf{q}_j)}{C_{D'}(\mathbf{q}_j)} \right) \\
    & \text{s.t. } (1) D' = D + \Delta D \quad (2)   |\Delta D| \leq K\notag\\\notag    
\end{align}}

\vspace{-1em}




Given the test workload $W_{Test}$  and a relation $R_x$ of $N$ tuples, denoted as $R_x = \{t_1, t_2, \ldots, t_N\}$.  We consider using $ t_i.\beta$ to represent the attacker's behavior on tuple $t_i$, where $t_i.\beta$ takes -1 to indicate delete tuple $t_i$ from the $R_{x}$, $t_i.\beta$ takes 0 to indicate no operation on tuple $t_i$, and $t_i.\beta$ takes $k$ to indicate duplicating  tuple $t_i$ by $k$ times and re-inserting them into the database.  Therefore, for the poisoned database $D'$, the cardinality of query $\textbf{q}_j$ on $D'$ equals to $C_{D'}(\textbf{q}_j) = C_D(\mathbf{q}_j) +  \sum_{i=1}^N t_i.\beta\times w_{ij} $. Based on the above, we reorganize the attack problem in Eq.~\ref{Eq001} into:

\textbf{Optimal Data-centric ACA (Optimal DACA) Problem}: Given testing workloads $W_{Test}$, the attacker needs to provide an attack strategy and maximize the following  Eq.~\ref{Eq003}: 

{\small
\begin{align}
\label{Eq003}
& { \text{Max}:\sum_{j=1}^M \text{max}\left(\frac{C_D(\mathbf{q}_j) + 1 + \sum_{i=1}^N t_i.\beta \times  w_{ij} }{C_D(\mathbf{q}_j)+1 },\frac{C_D(\mathbf{q}_j)+1}{C_D(\mathbf{q}_j) + 1 + \sum_{i=1}^N t_i.\beta \times  w_{ij} } \right) }\\
    & s.t.\sum_{i=1}^N |t_i.\beta| \leq K \quad t_i.\beta \in \{-1,0,1,\dots K\} \notag\\\notag    
\end{align}}


For the sake of brevity, we employ the abbreviation `optimal DACA' to represent `optimal Data-centric Algorithmic Complexity Attack' throughout the remainder of this paper. Meanwhile, we add 1 to both the numerator and denominator when calculating Qerror to avoid division by zero errors. This transformation is a commonly seen evaluation technique in many cardinality estimation methods and can be found in many open-sourced cardinality estimators' GitHub repositories e.g. line 4 in~\cite{alece_eval_utils}, line 21-22 in~\cite{price_eval_utils}. 

In summary, we converted the seemingly unsolvable black-box attack problem stated in Eq.~\ref{Eq001} into a constrained integer nonlinear programming problem, as detailed in Eq.~\ref{Eq003}. However, it remains uncertain whether the attacker can effectively solve Eq.~\ref{Eq003} in polynomial-time. Therefore, we will provide an analysis of intractability in the next section.








\subsection{\textcolor{black}{Intractability}}
\label{Sec.CC}


In this section, we give an intractability analysis of the optimal DACA problem defined in Eq.~\ref{Eq003}. We conclude that finding the optimal attack strategy is NP-Hard even when considering deletions alone. Additionally, when considering both insertions and deletions, finding an optimal attack strategy still remains NP-Hard. 




When only considering data deletions, the  problem is converted into the following optimization problem in Eq.~\ref{Opt.DelOnly}:

\vspace{-0.5em}
{\begin{align}
\label{Opt.DelOnly}
& \text{Max:}\sum_{j=1}^M \frac{C_D(\mathbf{q}_j)+1 }{C_D(\mathbf{q}_j) +1+ \sum_{i=1}^N t_i.\beta \times w_{ij} }  \\
    & s.t.\sum_{i=1}^N |t_i.\beta| \leq K\quad t_i.\beta\in \{-1,0\} \notag\\  \notag
\end{align}}
\vspace{-0.5em}

We now present a proof sketch that finds the optimal attack strategy in Eq.~\ref{Opt.DelOnly}, is NP-Hard. We start our polynomial-time reduction from the known  {Densest}-$K$-Subgraph  problem, defined as:\looseness=-1

\textbf{\uline{D}ensest}-$\uline{K}$\textbf{-\uline{S}ubgraph (DKS) problem:} 
Given a simple, undirected graph \( \mathbb{G} = (\mathbb{V}, \mathbb{E}) \) and an integer \( K \), find a subset of vertices \( S \subseteq \mathbb{V} \) such that: \( |S| \leq K\) and the subgraph \( \mathbb{G}[S] \) induced by \( S \) has the maximum  number of edges. 

\begin{figure}[htbp]
	\centering
	\includegraphics[width=7.5cm]{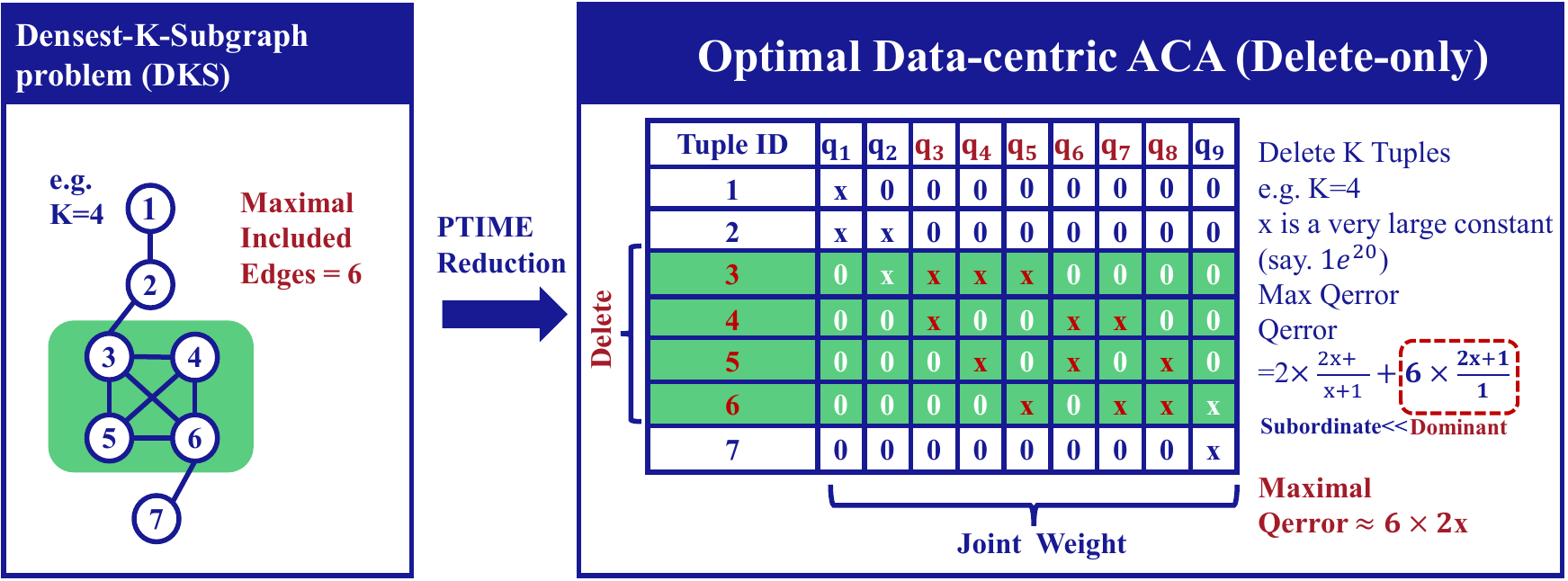}
         \vspace{-1em}
	\caption{Polynomial-time reduction example starting from Densest-K-Subgraph(DKS) problem.}
	\label{PNPHard}
 \vspace{-1em}
\end{figure}

The left subfigure in Figure~\ref{PNPHard} shows a DKS example when $K=4$. The DKS problem is a well-studied NP-Hard problem in graph theory~\cite{K-dense-Graph,DensestSubgraph_SODA,SODA17} and closely related to many applications in graph database community such as community search~\cite{k-CliqueDenset,YingliCS,zhou2024indepth}. We will next construct a polynomial-time reduction based on this problem.\looseness=-1

\begin{theorem}
    \label{thm:optimal-data-level-attack-np-hard}
    The optimal DACA problem defined in Eq.~\ref{Opt.DelOnly} when considering only deletions, is NP-Hard.
\end{theorem}

\begin{proof}
    {(Sketch)}We establish the NP-Hardness proof by providing a polynomial-time reduction from  the DKS  problem to the constrained maximization  in Eq~\ref{Opt.DelOnly} that finds $K$ deleting tuples to maximize the Qerror.

    For any DKS problem $P_{1}$, consisting of a graph \(\mathbb{G} = (\mathbb{V}, \mathbb{E}) \) and a number $K$, we can always construct a corresponding database instance and a group of testing queries to form the delete-only optimal DACA problem \( P_{2} \) in Eq~\ref{Opt.DelOnly} in time of $O(|\mathbb{V}| \cdot |\mathbb{E}| )$. The construction is as follows: (1)  Each vertex \( v_i \in \mathbb{V} \) corresponds  to a unique tuple $t_i$ in the relation \( R_x \). (2) Each edge \( e_{j} = \{v_{i1}, v_{i2}\} \in \mathbb{E} \) corresponds to a query $\textbf{q}_j$ involving only two tuples in $R_x$, $t_{i1}$ and $t_{i2}$. (3)  We assign each tuple contributes a common join weight \( \mathbf{x} \), where \( \mathbf{x} \) is a sufficiently large positive integer. 

    For the constructed delete-only DACA problem $P_{2}$, and a given  query $\textbf{q}_{j}$ consisting of two tuples. If one of its tuples is deleted, the contribution to the objective function is \( \Delta Qerr_1 = \frac{2\mathbf{x} + 1}{\mathbf{x} + 1} \), if two tuples are both deleted, the contribution is \( \Delta Qerr_2 = 2\mathbf{x} + 1 \).   When \( \mathbf{x} \) is sufficiently large, \(\Delta Qerr_1 \) approaches \( 2 \) and \( \Delta Qerr_2 \) approaches $2\mathbf{x}$ which is significantly larger: (\( \Delta Qerr_2 \gg \Delta Qerr_1 \)). Consequently, the maximization objective is predominantly influenced by \( \Delta Qerr_2 \). This implies that maximizing the objective function is effectively equivalent to maximizing \( 2\mathbf{x} \) times the number of edges in the \( K \)-densest subgraph of \( \mathbb{G} \). To better illustrate the above ideas, we use Example~\ref{EXNP} to show the reduction process.

     \begin{example}
     \label{EXNP}
 Figure~\ref{PNPHard} presents a reduction example. The left side depicts a DKS problem instance with \( K = 4 \), while the right side shows a corresponding optimal delete-only DACA problem for \( K = 4 \). For a graph with 7 vertices and 9 edges, we construct a database instance with a relation \( R_x \) containing 7 tuples and 9 queries. For instance, queries \( \textbf{q}_3 \), \( \textbf{q}_4 \), and \( \textbf{q}_5 \) encode connections between vertex 3 and vertices \{4, 5, 6\}. Each tuple contributes \( \mathbf{x} \) to its associated query's cardinality. When \( \mathbf{x} \) is sufficiently large (e.g., \( 10^{20} \)), deleting one tuple for a single query yields a Qerror contribution of approximately 2, while deleting two tuples results in \( 2\mathbf{x} \). Thus, the maximum Qerror arises when tuples from two queries are deleted. Moreover, the worst-case Qerror contribution scales with the number of internal edges in the \( K \)-vertex subgraph selected by DKS. In our example, the optimal DKS solution is a 4-vertex clique with 6 edges, leading to a maximum Qerror of \( 6 \times 2\mathbf{x} \) for the delete-only DACA.\looseness=-1
\end{example}

    Since the DKS problem is NP-Hard, and we have reduced it to the optimal delete-only DACA problem in polynomial-time, it follows that finding the optimal DACA problem when only considering deletions is NP-Hard.
\end{proof}

Based on the Theorem~\ref{thm:optimal-data-level-attack-np-hard}, we will demonstrate that considering both insertion and deletion simultaneously, the complete DACA problem is also NP-Hard.


\begin{theorem}
    \label{The.01}
    The optimal DACA problem defined in Eq.~\ref{Eq003} when considering both insertions and deletions, is NP-Hard.
\end{theorem}
\begin{proof}
	(Sketch): 
Our overall proof sketch is as follows: We extend the polynomial-time reduction constructed in Theorem~\ref{thm:optimal-data-level-attack-np-hard}. When the joint weight $\mathbf{x}$ is larger than $K \times M$, where \(K\) is the attack budget, and \(M\) is the number of queries, the benefits from $I$ insertion operations($ 2\leq I\leq K$) become negligible compared to delete two tuples within a given query $\textbf{q}_j$. This leads the attacker to abandon insertion operations and transform the case into a delete-only environment, and we can use the remainder of Theorem~\ref{thm:optimal-data-level-attack-np-hard}'s proof. Therefore, we establish a polynomial-time reduction from the DKS problem to the optimal DACA problem that considers both insertions and deletions, thereby proving that the optimal update strategy involving both insertions and deletions is NP-Hard. We provide  rigorous proof in our appendix~\cite{DACA_Appendix}.

\end{proof}


\textcolor{black}{ In summary,  the adversary cannot find an optimal solution to the DACA problem in polynomial-time. This theoretical result provides reassurance to database administrators, as identifying common data drift vulnerabilities that compromise all estimators is non-trivial. Such non-triviality implies certain robustness in these models, rendering worst-case vulnerabilities undiscoverable within polynomial-time. However, this does not guarantee inherent robustness for database systems using learned estimators. As demonstrated in the following section, attackers can design polynomial-time approximation algorithms with provable guarantees to seek near-optimal worst-case situations.}



\section{Near optimal Attack Deployment}
\label{Sec.DeployATK}
In the previous section, we obtained some bad news for the adversary that no polynomial-time algorithm can find optimal data drifts that brings the worst case to all estimators. Fortunately, the optimal DACA problem we are studying possesses certain favorable properties, which allow for the existence of polynomial-time algorithms capable of providing exact or approximately optimal attack strategies under specific conditions. In \S~\ref{sec.PropertyA}, we will analyze these favorable properties. In \S~\ref{sec.AUAF}, we will utilize these special properties to design an efficient approximate algorithm. In \S~\ref{sec.AAS}, we will analyze the algorithm in \S~\ref{sec.AUAF}.\looseness=-1



\subsection{Analysis of the DACA's Objective Function}
\label{sec.PropertyA}

In this section, we will analyze the nature of DACA's objective function under specific conditions to derive certain insights.




To facilitate the presentation of symbols in the following discussion, we denote the Qerror induced by the \( j \)-th query, considering only deletions, as: $ 
\mathbb{Q}_j(X) = \frac{C_D(\mathbf{q}_j) + 1}{C_D(\mathbf{q}_j) + 1 + \sum_{t_i\in X} t_i.\beta\times w_{ij}}$. $X$ is the deleted tuples within the relation $R_x$, having $X\subseteq R_x, \forall t_i\in X, t_i.\beta=-1$. Therefore, the total Qerror when only deletion is considered is $\mathbb{Q}(X) = \sum_{j=1}^M \mathbb{Q}_j(X)$. Based on these definitions, we have Theorem~\ref{The.51}\looseness=-1

\begin{theorem}
    \label{The.51}
    When only deletion operations are considered, total Qerror $ \mathbb{Q}(X)$ of the optimal DACA problem satisfies the supermodular property, that is: 
    $\forall A,B \subseteq R_{x}, \mathbb{Q}(A \cup B) + \mathbb{Q}(A \cap B) \geq \mathbb{Q}(A) + \mathbb{Q}(B).$
\end{theorem}

\begin{proof}
    (Sketch)We aim to demonstrate that the objective
   {\small \[
    \mathbb{Q}(X) = \sum_{j=1}^M \mathbb{Q}_j(X) = \sum_{j=1}^M \frac{C_D(\mathbf{q}_j) + 1}{C_D(\mathbf{q}_j) + 1 + \sum_{t_i\in X} t_i.\beta\times w_{ij}}
    \]}
    satisfies the supermodular property. Specifically, for any two sets \( A \) and \( B \) within $R_x$, we need to prove that $\forall A,B \subseteq R_{x}, \mathbb{Q}(A \cup B) + \mathbb{Q}(A \cap B) \geq \mathbb{Q}(A) + \mathbb{Q}(B).$

    First, consider each component function \( \mathbb{Q}_j(X) \). We aim to show that \( \mathbb{Q}_j(X) \) is supermodular. We denote:
    {\small\[
    w'_{ij} = \frac{w_{ij}}{1 + C_D(\mathbf{q}_j )} ,\quad S_j(X) = \sum_{t_i\in X} t_i.\beta \times w'_{ij}
    \]}
    Then, the expression:
    \[
    \mathbb{Q}_j(A \cup B) + \mathbb{Q}_j(A \cap B) - \mathbb{Q}_j(A) - \mathbb{Q}_j(B)
    \]

    can be reorganized into the following form,  we provide the detailed derivation in the appendix~\cite{DACA_Appendix}.
    
  {\small  \[
  \quad \frac{(2+S_j(A)+S_j(B))(S_j({A \setminus B}))(S_j({B \setminus A}))}{(1 + S_j({A \cup B}))(1 + S_j({A \cap B}))(1 + S_j(A))(1 + S_j(B))}
    \]}


    Observe that given $\forall t_i\in X, t_i.\beta=-1, w'_{ij}\geq 0 $, we have $S_j({X})$ is negative. We deduced that the two factors of the numerator, $S_j({A \setminus B})$ and $S_j({B \setminus A})$, are both less than or equal to 0. Thus, their product is greater than or equal to 0. 
    Meanwhile, $\forall X\subseteq R_x,  (1+S_j(X)) \geq (1+S_j(R_x))= \frac{1}{1+C_D(\mathbf{q}_j )}  > 0$.  
    Therefore, each term in the denominator is strictly positive, and $(2+S_j(A)+S_j(B))$ is strictly positive. In conclusion, within the above fractional, the numerator is greater than or equal to 0, while the denominator is strictly greater than 0. We have:\looseness=-1
    
    \[
    \mathbb{Q}_j(A \cup B) + \mathbb{Q}_j(A \cap B) - \mathbb{Q}_j(A) - \mathbb{Q}_j(B) \geq 0,
    \]
    
    which confirms that \( \mathbb{Q}_j(X) \) is supermodular.   Since \( \mathbb{Q}(X) \) is the sum of supermodular functions \( \mathbb{Q}_j(X) \), it inherits the supermodular property. Formally,
    \begin{align*}
    & \mathbb{Q}(A \cup B) + \mathbb{Q}(A \cap B) - \mathbb{Q}(A) - \mathbb{Q}(B) \\
    =& \sum_{j=1}^M \left[ \mathbb{Q}_j(A \cup B) + \mathbb{Q}_j(A \cap B) - \mathbb{Q}_j(A) - \mathbb{Q}_j(B) \right]
    \geq 0.
    \end{align*}
    Hence, the objective \( \mathbb{Q}(X) \) satisfies the supermodular property.
\end{proof}

Next, we consider the scenario where only insertions are considered, the attack problem discussed in Eq.~\ref{Eq003} can be much simpler. Similar to Theorem~\ref{The.51}, we define the Qerror induced by the \( j \)-th query, considering only insertions, as: $ 
\mathbb{Q}'_j(X) = \frac{C_D(\mathbf{q}_j) + 1+ \sum_{t_i\in X} t_i.\beta\times w_{ij}}{C_D(\mathbf{q}_j) + 1 } $. And $X$ is the set of tuples that are repeatedly inserted into $R_x$, having $X\subseteq R_x, \forall t_i\in X, t_i.\beta=1$. Therefore, the total Qerror when only insertion is considered is $\mathbb{Q}'(X) = \sum_{j=1}^M \mathbb{Q}'_j(X)$. Based on these definitions, we have Theorem~\ref{The.03}:


\begin{theorem}
    \label{The.03}
    When only considering insertions, the optimal data-level attack problem satisfies the modular property, i.e., $\forall A,B \subseteq R_x, \mathbb{Q}'(A \cup B) + \mathbb{Q}'(A \cap B) = \mathbb{Q}'(A) + \mathbb{Q}'(B).$
    
\end{theorem}
\begin{proof}
    We leave the detailed proof in our appendix~\cite{DACA_Appendix}.\looseness=-1
\end{proof}


The above theorems indicate that although the optimal DACA problem in Eq.~\ref{Eq003} is NP-Hard, the objective function possesses special properties when the attack operation types are limited. We will utilize this characteristic to design an effective approximate attack algorithm and analyze the attack's effectiveness. 

\subsection{Attack Strategy Generation}
\label{sec.AUAF}
In this section, we will utilize the characteristics analyzed in the previous section to design a corresponding algorithm to maximize Eq.~\ref{Eq003}. We first present the attack strategy  in Algorithm~\ref{alg.PATK}:

\begin{algorithm}[htb]
	\caption{Approximate solution to optimal DACA.}
	\label{alg.PATK}
	\begin{algorithmic}[1] 
		\Require
		{ Table $R_x$; Testing queries $W_{Test}$; Testing queries' results $\mathcal{R}es$; Budget $K$ ;  }
		\Ensure Modification set $\Delta D$ with at most $K$ modifications;
 \State $w = getJointWeight (\mathcal{R}es,R_x)$  \Comment{ Obtaining joint weight}
\State $\Delta D = \phi; Mask = \textbf{0}_{N}$
 \While{$ |\Delta D| \leq K $} 
     
 \State $bestOp = \phi; bestVal = 0;$
 \For{ {\small $(t_i\in R_x) \land (Mask[i] =0) $}  } \Comment{ {\footnotesize Get $R_x$'s non-deleted tuples} }
 
 \State $gainD = \sum\limits_{\textbf{q}_j \in W_{Test}} Q(C_{D+\Delta D}(\textbf{q}_j),C_{D+\Delta D}(\textbf{q}_j)-w_{ij})$;

 \If{$ gainD \geq bestVal$}
 \State{$bestVal=  gainD$ }
 \State{$bestOp = -t_i$}
 \EndIf
 
 \State $gainI= \sum\limits_{\textbf{q}_j \in W_{Test}} Q(C_{D+\Delta D}(\textbf{q}_j),C_{D+\Delta D}(\textbf{q}_j)+w_{ij})$;
 \If{$ gainI \geq bestVal$}
 \State{$bestVal=  gainI$ }
 \State{$bestOp = t_i$}
 \EndIf


 \EndFor
 

 \If{$bestOp = \phi$}
\State $break$
 \EndIf
 \If{$bestOp = -t_{\delta}$} \Comment{Mask the deleted tuple}
\State $Mask[\delta]=1$
 \EndIf
 
 \State $\Delta D.append(bestOp);$
 \EndWhile
		\\
		\Return {$\Delta D$}; 
	\end{algorithmic}
\end{algorithm}


The general idea of the Algorithm~\ref{alg.PATK} is as follows. In line~1, the attacker performs a group-by operation on the result $\mathcal{R}es$ of the testing queries $W_{Test}$ over the relation \( R_x \) and calculates the joint weight of each tuple in \( R_x \) with respect to each query join. For a given  query $\textbf{q}_j$  and its materialized result in $\mathbbm{S}(\textbf{q}_j|D)$, this operation can be executed via the following SQL:
\begin{align*} 
\label{EQ.Group By}
& \texttt{SELECT } R_x.PK, \texttt{COUNT(*) } \texttt{FROM } \mathbbm{S}(\textbf{q}_j|D)  \texttt{ GROUP BY }  R_x.PK;
\end{align*}

where \textit{$R_{x}.PK$} is the primary key of the relation \textit{$R_{x}$}. 

Lines 2 and 4 of the algorithm handle the initialization of variables. Lines 3-17 employ a greedy approach to sequentially select operations that maximize the current Qerror. Specifically, lines 5-13 iterate through the modified relation table one by one, computing the weights for both deletion and duplicate insertion for each tuple. In each iteration, the algorithm greedily selects the operation that maximizes the local Qerror. Lines 16-17 temporarily save the deleted tuples to prevent duplicate computation on already deleted tuples.\looseness=-1


The aforementioned algorithm guarantees termination within polynomial-time. Specifically, the time complexity of the algorithm is \( {O}(  |\mathcal{R}es| + N \times K \times M) \), where: $|\mathcal{R}es|$ is the overhead of the group by operation in line~1 and is linear to the size of the result set $\mathcal{R}es$, 
 \( N \) is the cardinality of the relation \( R_x \),
 \( K \) is the budget allocated to the attacker for data insertion, and \( M \) is the size of the test query.\looseness=-1

\subsection{Analysis on the Attack Strategy }
\label{sec.AAS}


In this section, we evaluate the effectiveness of our attack strategy outlined in Algorithm~\ref{alg.PATK}. We demonstrate that the strategy can deliver optimal or near-optimal results in scenarios limited to either insertions or deletions. Additionally, we present two corollaries to showcase its performance in more general situations. Initially, we establish that when operations are confined to deletions, Algorithm~\ref{alg.PATK} achieves a \(1-\kappa\) approximation ratio.


\begin{theorem}
    \label{The.04}
    In the scenario where only deletions are considered, let \( \mathbb{Q}(O) \) denote the optimal attack result. Algorithm~\ref{alg.PATK} can achieve an approximation of \( (1-\kappa) \). This is, the output \( \Delta D \) of the Algorithm~\ref{alg.PATK} possesses the property that $\mathbb{Q}(\Delta D) \geq (1-\kappa) \mathbb{Q}(O)$, where $\kappa = 1 - \min\limits_{t \in R_x} \min\limits_{A,B\subseteq {R_x-t} }\frac{\mathbb{Q}(A+t)-\mathbb{Q}(A)}{\mathbb{Q}(B+t)-\mathbb{Q}(B)}.$
\end{theorem}
\begin{proof}
    (Sketch).
    

    Based on the definition of \(\kappa\), we obtain \(\forall A, B \subseteq R_x, 1 - \kappa \leq \min\limits_{t \in R_x} \frac{\mathbb{Q}(A+t)-\mathbb{Q}(A)}{\mathbb{Q}(B+t)-\mathbb{Q}(B)}\). Furthermore, we let the set \(\Delta D_i\)  denote the set chosen by Algorithm~\ref{alg.PATK} at the \(i\)-th step, and we let \(\Delta D_i\) to substitute \(A\). Similarly, we can arbitrarily choose a subset permutation of the optimal solution $O$ and let a subset \(O_i\) with size \(i\) from the permutation to substitute \(B\). We also define \(t_{i1} = O_{i+1} - O_i\) and \(t_{i2}=\Delta D_{i+1}-\Delta D_i\). By definition, we can derive that:
{\small
\[
1 - \kappa \leq \frac{\mathbb{Q}(t_{i1} + \Delta D_i) - \mathbb{Q}(\Delta D_i)}{\mathbb{Q}(t_{i1} + O_i) -\mathbb{Q}(O_i)}
\]
}
Thus, we obtain:
  \vspace{-0.5em}
{\small\begin{equation}
\label{EQT}
\left(\mathbb{Q}(t_{i1} + O_i) -\mathbb{Q}(O_i)\right) \times (1 - \kappa) \leq \mathbb{Q}(t_{i1} + \Delta D_i) - \mathbb{Q}(\Delta D_i)    
\end{equation}}

Given that the outer loop of the Algorithm~\ref{alg.PATK} runs for \(K\) steps, we can apply the aforementioned operation at each step of the  process. Consequently, we have:
  \vspace{-0.5em}
{\small\[
(1 - \kappa) \mathbb{Q}(O) = (1 - \kappa) \mathbb{Q}(\emptyset) + (1 - \kappa) \sum_{i=1}^K \mathbb{Q}(O_{i}+t_{i1})-\mathbb{Q}(O_{i}).
\]}
Since \(\mathbb{Q}(\emptyset) \geq 0\) and combining Eq.~\ref{EQT}, we can conclude:
{\small
\[
(1 - \kappa) \mathbb{Q}(O) \leq \mathbb{Q}(\emptyset) + \sum_{i=1}^K  \mathbb{Q}(t_{i1} + \Delta D_i) - \mathbb{Q}(\Delta D_i)  
\]
}

Finally, based on lines~6-9 of Algorithm~\ref{alg.PATK} at each step, we have $\left( \mathbb{Q}(t_{i2} + \Delta D_i) - \mathbb{Q}(\Delta D_i) \right) \geq \left( \mathbb{Q}(t_{i1} + \Delta D_i) - \mathbb{Q}(\Delta D_i)\right) $. Therefore:
  \vspace{-0.4em}
{\small
\begin{align*}
    (1 - \kappa) \mathbb{Q}(O) &\leq \mathbb{Q}(\emptyset) + \sum_{i=1}^K \left( \mathbb{Q}(t_{i1} + \Delta D_i) - \mathbb{Q}(\Delta D_i) \right) \\
    &\leq \mathbb{Q}(\emptyset) + \sum_{i=1}^K \left( \mathbb{Q}(t_{i2} + \Delta D_i) - \mathbb{Q}(\Delta D_i) \right) \\
    &= \mathbb{Q}(\Delta D_K).
\end{align*}}

This means that:
{\small\[
(1 - \kappa)\times \mathbb{Q}(O) \leq \mathbb{Q}(\Delta D_K).
\]}
Thus, we  prove that the approximation ratio of  algorithm is \(1 - \kappa\).

\end{proof}






After considering the scenario where only deletions are made, we will continue to analyze our algorithm's performance in the context of insertion-only scenarios. Fortunately, we conclude that, when considering only insertions, our algorithm is capable of achieving the optimal solution constituted solely by insertions.

\begin{theorem}
    \label{The.06}
    In the case of considering only insertions, the above greedy approach can reach an optimal insertion result.
\end{theorem}
\begin{proof}
    (Sketch). 
    \textbf{Base Case ($K=1$):} When $K=1$, the algorithm iterates through all possible values and identifies the tuple that maximizes the gain function $G$ for insertion. Therefore, the algorithm is optimal at this initial step.  
    \textbf{Inductive Step ($K \geq 2$):} Assume that for $K-1$, the algorithm achieves optimality. We need to show that under this assumption, the algorithm also achieves optimality for $K$. Suppose that $\Delta D_{K-1}$ achieves the maximal Qerror and inserts $K-1$ tuples. Let the optimal insertion attack strategy select tuple $t_a$ at the $K$-th step, while our Algorithm~\ref{alg.PATK} selects $t_b$. According to lines~10-13 of Algorithm~\ref{alg.PATK}, the  following inequality holds:
{\small    \[
    \sum_{j=1}^M \frac{w_{bj}}{C_D(\mathbf{q}_j) + 1} \geq \sum_{j=1}^M \frac{w_{aj}}{C_D(\mathbf{q}_j) + 1}
    \]}
    This implies that $\mathbb{Q}'\left( t_b+ \Delta D_{K-1} \right) \geq \mathbb{Q}'\left( t_a+ \Delta D_{K-1} \right).$ Thus, if that optimality is achieved at step $K-1$, using lines~10-13 of the Algorithm~\ref{alg.PATK} ensures that the algorithm remains optimal for step $K$.

\end{proof}


\vspace{-1em}

Although the two aforementioned theorems are only preliminary considering insert-only or delete-only optimal DACA scenarios. We will now demonstrate, through the following corollaries, that in certain special cases, even when considering both insertions and deletions simultaneously, our algorithm can still achieve good approximation performance. The proof of these corollaries can be found in the appendix~\cite{DACA_Appendix}.

\textbf{Corollary1:} If the attacker can pre-determine whether each query in the testing workload is insertion-dominated or deletion-dominated, this mathematically removes the inner maximization layer of the Qerror summation in Eq.~\ref{Eq003}. Under this condition, $\Delta D$ of Algorithm~\ref{alg.PATK} retains the following property: $\mathbb{Q}(\Delta D) \geq (1-\kappa) \mathbb{Q}(O)$.


\textbf{Corollary 2:} If the  queries meet the following conditions:  $M > K$,  and for all queries \(\textbf{q}_j\), we have $
\text{count}(w_{i,j} \neq 0) \leq 2$ and $C_D(\textbf{q}_j)> K\times M $. Then, Algorithm~\ref{alg.PATK} still possesses the following property: $\mathbb{Q}(\Delta D) \geq (1-\kappa) \mathbb{Q}(O)$.





\textcolor{black}{The theorem and corollary demonstrate that while computing the optimal DACA is NP-Hard, Algorithm~\ref{alg.PATK} offers a polynomial-time \((1-\kappa)\)-approximation strategy under practical constraints. The analysis presented allows us to approximate the worst case of learned estimators within polynomial-time, prior to the occurrence of catastrophic estimation errors. This insight enables database administrators to effectively quantify potential weaknesses in  estimators, providing a foundation for robust query optimization strategies.}\looseness=-1


\section{Experiment}
\label{sec.exp}

In this section, we provide an experimental analysis of the effectiveness and scalability  of the data-centric algorithmic complexity attacks on learned estimators and their potential consequences. We will use our experiment results to answer the following questions:\looseness=-1

1. \textbf{Effectiveness:} Can the proposed DACA techniques effectively compromise the performance of data-driven, query-driven, and hybrid estimators? (\S~\ref{Declie.Performance})

2. \textbf{Scalability:}  How do variations in data size, query scale, and the selection of attacked tables influence the efficiency and effectiveness of the proposed DACA technique? (\S~\ref{Sec.ScaEva})

3. \textbf{{Consequence \& Defense:}} How would the compromised estimator  mislead the query optimizer into selecting suboptimal query plans?  What are the characteristics of such plans? Do we have countermeasures defend against DACA and enhance the worst-case performance for learned estimators? (\S~\ref{Sec.Consequence})





\subsection{Experimental Setup}

\begin{figure*}[htbp]

 \subfigure[Mean of Qerror on STATS-CEB.]{
		\label{Fig.STATS_QErr}
		\includegraphics[width=1.8\columnwidth]{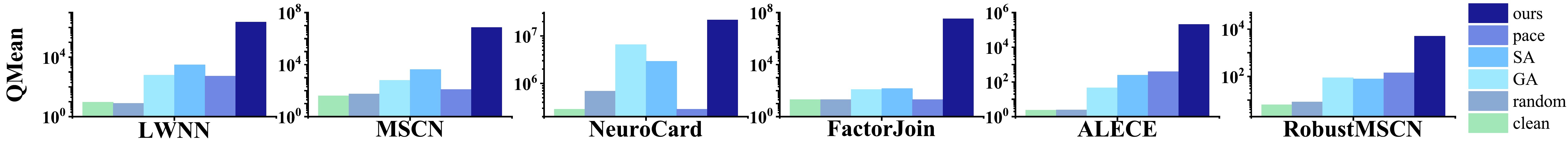}
	}  
    \vspace{-1.0em}    
	
 \subfigure[ Mean of Qerror on IMDB-JOB. ]{
		\label{Fig.IMDB_QErr}
		\includegraphics[width=1.8\columnwidth]{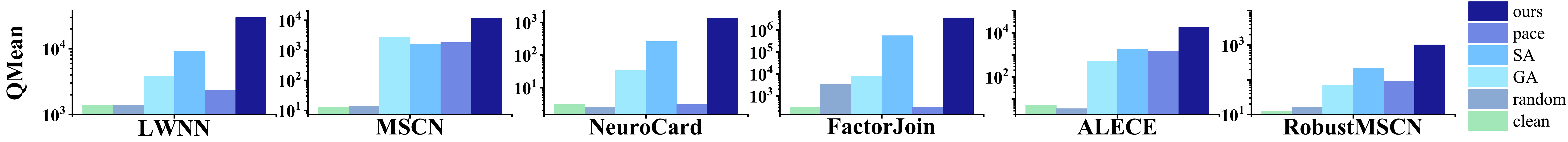}
	}

    \vspace{-1.0em}    


 \subfigure[End-to-end query performance on STATS-CEB.]{
		\label{Fig.IMDB_PerQueryE2E}
		\includegraphics[width=1.8\columnwidth]{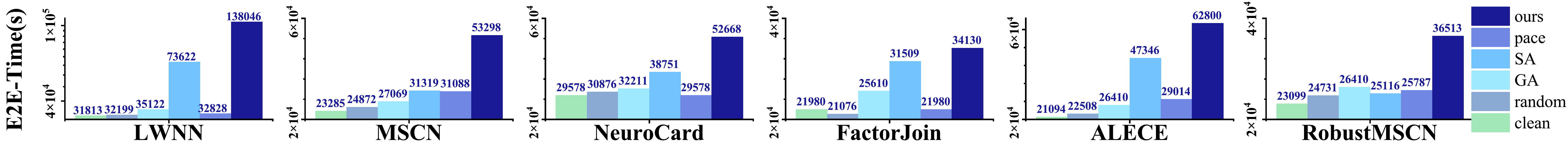}
	}

    \vspace{-1.0em}    

 \subfigure[End-to-end query performance on  IMDB-JOB.]{
		\label{Fig.STATS_PerQueryE2E}
		\includegraphics[width=1.8\columnwidth]{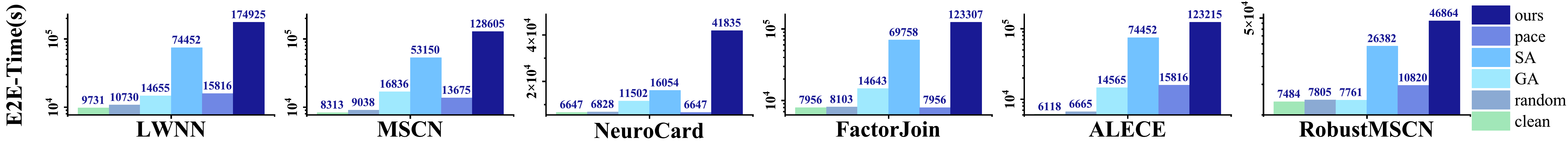}
	}  
    \vspace{-1.0em}    

	\caption{ Mean of Qerrors and  end-to-end(E2E) latency degradation per query on STATS/IMDB database.}
	\label{Exp.Mean}
	\vspace{-1.0em}    
\end{figure*}


\textbf{\uline{Datasets}:} To evaluate the performance of attack methods in multi-table scenarios, we selected the widely adopted benchmarks IMDB-JOB and STATS-CEB for assessment, as they are extensively used in the evaluation of multi-table cardinality estimators~\cite{ASM,FactorJoin,ALECE,PRICE,RobustMSCN,DBMSCE}. The IMDB-JOB benchmark~\cite{JOB_Light} is based on the IMDB movie database~\cite{IMDB} and comprises six relations (\texttt{cast\_info, movie\_info, movie\_companies, movie\_keyword, movie\_info\_idx,  title}) and fourteen attributes, totaling 62,118,470 records. For the test workload, we selected the open-source JOB-light workload~\cite{JOBLightAndSUBQ}, which consists of 70 real queries and 696 subqueries. Additionally, STATS-CEB~\cite{DBMSCE}  is a collection of user-generated anonymized content from stack exchange network~\cite{StackExchange}  including eight relations (\texttt{users, posts, postLinks, postHistory, comments, votes, badges,  tags}) and 43 attributes, comprising a total of 1,029,842 records. For the test workload, we selected the open-source workload version~\cite{STATSAndSUBQ} consisting of 146 queries and 2,603 subqueries. To train the query-driven model on these databases, we followed the approach of Li et al.~\cite{ALECE}, generating 2,000 corresponding training queries and their respective subqueries for training.\looseness=-1

\textbf{\uline{CE models}:} Based on various studies~\cite{ALECE,FactorJoin,RobustMSCN},  we selected the following most competitive and representative learned estimators covering the paradigms of data-driven, query-driven, and hybrid:

(1) \textit{LWNN (Query-driven)}~\cite{LWNN}:  uses MLPs to map query representations to cardinalities. We follow the instructions in~\cite{DBMSCE} to extend LWNN to support joins.

(2) \textit{MSCN (Query-driven)}~\cite{MSCN_Kipf2018LearnedCE}: the most famous learned query-driven method based on a multiset convolutional network model.

(3) \textit{NeuroCard (Data-driven)}~\cite{nec}: learns a single deep auto-regressive model for the joint distribution of all tables in the database and estimates the cardinalities using the learned distribution. Following the settings in~\cite{nec}, we set the sampling size of the NeuroCard to 8,000.\looseness=-1

(4) \textit{Factorjoin (Data-driven)}~\cite{FactorJoin}: integrates classical join-histogram methods and learned Bayes Network into a factor graph.

(5) \textit{ALECE (Hybrid)}~\cite{ALECE}: uses a transformer to learn the mapping from query representation and histogram features to cardinality.  

(6) \textit{RobustMSCN (Hybrid)}~\cite{RobustMSCN}: is an improvement to the basic MSCN method~\cite{MSCN_Kipf2018LearnedCE}, aims at improving the robustness of the naive MSCN by employing masked training and incorporating data-leveled features such as PG estimates and join bitmaps.  

(7) \textit{Oracle}: is the oracle $E_O$ which is capable of utilizing all information from the poisoned database \(D'\) to obtain query $\textbf{q}$'s true cardinality  \(C_{D'}(\textbf{q})\) within \(D'\).

\textbf{\uline{Baselines}:} We compare DACA with the estimators before any attacks (Clean). Meanwhile, we compare the four attack baselines as follows, we selected these baselines because they have either been used to reduce the performance demands on learned cardinality estimators or have been employed to solve similar NP-Hard database optimization problems~\cite{Yannis1,Yannis2,PACE}.


(1)  \textit{PACE}~\cite{PACE}: is the State-Of-The-Art query-centric attack method on learned query-driven estimators.  PACE compromises query-driven estimators by poisoning historical workloads.

(2) \textit{Random Generation (Random)}: randomly selects tuples from the training database to delete or duplicate.

(3) \textit{Simulated Annealing (SA)}~\cite{SA}: constructs attack strategies via a probabilistic hill-climbing algorithm and maximizes Eq.~\ref{Eq003}.

(4) \textit{Genetic Algorithm (GA)}~\cite{GA}: treats a set of potential attack strategies as a gene, by randomly selecting several sets of possible attack strategies and using them as a population for genetic crossover and mutation, therefore  maximizing Eq.~\ref{Eq003}.




\textbf{\uline{Experimental Settings}:}   \textit{Estimators Settings}: For the cardinality estimators to be attacked, we train them on the poisoned database $D'$ and test them on the clean database $D$. That is to say, the cardinality labels for the Qerror evaluation and the data used to evaluate the actual latency are both derived from the clean database $D$. For hybrid methods, we provide the data features of the clean database during testing. Specifically, we supply ALECE with histogram representations from the clean database $D$, and provide RobustMSCN with PG estimates and joint sampling features from the clean database. 
\textit{Attack Settings}: In \S~\ref{Declie.Performance}, we target the \texttt{title} table of the IMDB-JOB database and the \texttt{posts} table of the STATS-CEB database. We set the data-centric attack budget to 20\% of the attacked table. This means that in the IMDB-JOB database, the attacker performs $5 \times 10^{5}$ tuple leveled drifts, altering a data scale of $\frac{5 \times 10^{5}}{6.2 \times 10^{7}} = 0.008$. Similarly, in the STATS-CEB database, the attacker conducts $1.8 \times 10^{4}$ update operations, changing a data scale of $\frac{1.8 \times 10^{4}}{1 \times 10^{6}} = 0.012$.\looseness=-1


\textbf{\uline{Metrics}:} We use the metrics defined in \S~\ref{sec.threatModel} to evaluate the attack's effectiveness. We use Qerror to measure the degradation in estimation accuracy to evaluate the quality of the generated query plans, and end-to-end (E2E) time to assess the impact of estimation on the end-to-end execution of queries. We utilize the framework developed in~\cite{DBMSCE} to inject the cardinalities predicted by the estimator into Postgres and evaluate the end-to-end execution time.\looseness=-1

\textbf{\uline{Environments}:} Our end-to-end experiments were conducted on an individual Huawei Cloud server with 4 Intel(R) Xeon(R) Platinum 8378A CPUs and 32GB RAM and 1T SSD. Apart from that, all remaining experiments were run on a server with 4 RTX A6000 GPUs, 40 Intel(R) Xeon(R) Silver 4210R CPUs, and 504GB RAM. 

\vspace{-1em}
\subsection{Decline of the Estimator's Performance}
\label{Declie.Performance}

\textbf{Average Decline of the Estimator's Qerror:} The decline in the estimators' average Qerror serves as an indicator of the overall attack efficacy of a given strategy~\cite{PACE}. This metric is directly aligned with the primary maximization objective in both PACE~\cite{PACE} and our proposed formulation in Eq.~\ref{Eq001}, thereby reflecting the attack's effectiveness. Figure~\ref{Fig.STATS_QErr} and~\ref{Fig.IMDB_QErr} illustrate the average Qerror of each estimator before (Clean) and after the attack on the IMDB-JOB and STATS-CEB datasets. For models such as LWNN, MSCN, and ALECE, the attack efficacy ordering is Ours > SA > PACE > GA > Random. On average, our approach outperforms the four baselines by factors of $6.85\times$, $10.47\times$, $15.27\times$, and $1876\times$, respectively. For the remaining data-driven and hybrid models, the ordering is Ours > SA > GA > PACE > Random, with our method surpassing the baselines by $4.92\times$, $26.9\times$, $225\times$, and $293\times$, respectively. Our analysis reveals that PACE's attack effectiveness is limited to models reliant on historical query data (e.g., LWNN, MSCN, and ALECE). This stems from PACE's approach of poisoning historical queries, which disrupts the mapping between query representations and cardinalities or corrupts the histogram-based lookup tables used by these estimators. In contrast, methods such as NeuroCard   and RobustMSCN  leverage data-level knowledge independent of historical workloads, enhancing their robustness against such poisoning attacks. Unlike PACE, our data-level attack strategy contaminates the underlying data distributions and correlations, thereby affecting both data-driven and hybrid estimation methods. Furthermore, our proposed attack strategy demonstrates superior efficacy compared to SA, GA, and Random attacks. This observation aligns with the conclusions drawn in \S~\ref{sec.AAS}, where despite the NP-Hard nature of optimizing Eq.~\ref{Eq003}, Algorithm~\ref{alg.PATK} effectively identifies near-optimal solutions.\looseness=-1


\begin{table*}[h]
 \scriptsize
\caption{Percentile Qerrors on STATS-CEB and IMDB-JOB.}
\label{SuperBigTab}
 \scalebox{0.85}{

 \begin{tabular}{c|c|c|lllll|lllll}
\hline

      &  &   & \multicolumn{5}{c|}{\sf IMDB-JOB QERROR} & \multicolumn{5}{c}{\sf STATS-CEB QERROR}\\ \cline{4-13} 

\multirow{-2}{*}{\sf PARADIGM}& \multirow{-2}{*}{ \sf CE METHOD} & \multirow{-2}{*}{\sf ATTACK METHOD}   & \multicolumn{1}{l}{ \sf 50\%} & \multicolumn{1}{l}{\sf 90\%} & \multicolumn{1}{l}{\sf 95\%} & \multicolumn{1}{l}{\sf 99\%} & \sf MAX     & \multicolumn{1}{l}{ \sf 50\%} & \multicolumn{1}{l}{\sf 90\%} & \multicolumn{1}{l}{\sf 95\%} & \multicolumn{1}{l}{\sf 99\%} & \sf MAX     \\ \hline
\multirow{12}{*}{Query-Driven}
& \multirow{6}{*}{LWNN}& Clean& $2.661$& $65.04$& $117.6$& $1567$& $5.67 \cdot 10^{4}$
    & $4.762$   & $25.98$ & $47.18$ & $1182$& $2.66\cdot10^{4}$ \\  
  && Random& $2.271$ & $64.52$ & $118.2$& $1595$ & $3.81\cdot 10^{4}$ 
    & $4.489$   & $20.37$& $32.10$ & $1505$ & $3.51\cdot 10^4$   \\  
  && GA   & $48.61$& $1105$ & $5570$ & $8.41 \cdot 10^4$& $9.60 \cdot 10^4$  
    & $21.84$   & $377.7$ & $1167$ & $9173$  &  $5.52\cdot 10^4$  \\  
  && SA   & $184.1$ & $2.05\cdot10 ^4$ & $4.21\cdot10^4$ & $1.28\cdot10^5$ & $1.91\cdot10^5$   
  & $39.87$   & $1065$ & $3847$ & $4.65 \cdot 10^4$ & $3.67\cdot 10^6$\\  
  && PACE & $27.54$& $1359$ & $6542$& $4.79\cdot10^4$ & $9.59 \cdot 10^4$  & $26.44$   & $654.8 $& $2486 $& $1.36 \cdot10^4$ & $7.23\cdot 10^7$ \\  
  && \textbf{Ours}    & $\mathbf{350.3}$ & $\mathbf{{3.29}\cdot10^4}$& $\mathbf{9.12\cdot10^4}$ & $\mathbf{1.51\cdot10^5}$ & $\mathbf{9.35\cdot10^8}$   & $\mathbf{532.8}$   & $\mathbf{2.43\cdot10^4}$& $\mathbf{8.35\cdot10^4}$& $\mathbf{1.14\cdot10^6}$ & $\mathbf{1.14\cdot10^8}$ \\ \cline{2-13} 
  
  & \multirow{6}{*}{MSCN}& Clean& $2.076$& $13.92$ & $53.68$ & $199.3 $& $1653$  
  & $2.235 $  & $99.10$ & $179.1 $& $505.1$ & $2689 $  \\  
  && Random& $2.085$ & $20.14$ & $59.14$ & $178.1$ & $2116$  
  & $2.724$   & $161.8$ & $298.8$& $732.2$& $4342$  \\  
  && GA   & $18.58$& $443.2$ & $2111$ & $2.13\cdot10^4$& $1.12\cdot10^6$
  & $19.71$    & $397.2$ & $1257$ & $1.09\cdot10^4$ & $5.03\cdot10^4$  \\  
  && SA   & $30.07$& $2198$& $6911$& $2.51\cdot10^4$& $1.71\cdot10^5$  
  & $38.31$   & $3731$ & $1.58\cdot10^4$ & $1.05\cdot10^5$& $1.24\cdot10^5$  \\  
  && PACE & $54.60$& $569.1$ & $1437$ & $4.46\cdot10^4$ & $2.11\cdot10^5$  & $10.01$   &$ 117.6 $& $209.2$& $1488$& $9724$ \\  
  && \textbf{Ours}    & $\textbf{60.23}$ 
  & $\textbf{6201}$  &  $\mathbf{3.04 \cdot 10^4} $ &
  $ \mathbf{1.56\cdot 10^5}$ &   {$\mathbf{1.46 \cdot 10^6}$ }& {$\mathbf{184.3}$}   & {$\mathbf{4.63\cdot 10^6} $}& {$\mathbf{1.16 \cdot 10^7}$} & {$\mathbf{4.74 \cdot 10^7}$} & {$\mathbf{8.08\cdot 10^8}$} \\ \hline
  
\multirow{12}{*}{Data-Driven}  & \multirow{6}{*}{NeuroCard}  
   & Clean& $1.416$& $3.312$ & $8.32$ & $18.39$ & $21.51$   & $2.621$   & $1089$ & $6643$ & $4.95\cdot10^4$ & $1.71\cdot10^6$ \\  
  && Random& $1.869$ & $4.632$ & $8.533$ & $25.97$ & $27.03$  & $2.986$   & $1092$ & $6426$ & $4.95\cdot10^4$ & $4.85\cdot10^6$\\  
  && GA   & $1.884$& $11.72$& $56.56$ & $830.9$ & $971.6$  & $4.09$   & $1073$ & $6690$ & $6.6\cdot10^5$ & $1.04\cdot10^6$  \\  
  && SA   & $39.58$& $484$& $1031$ & $3626$& $6618$ & $3.967$   & $1181$ & $7211$ & $4.95\cdot10^5$ & $1.04\cdot10^7$ \\  
  && PACE& $1.416$& $3.312$ & $8.32$ & $18.39$ & $21.51$   & $2.621$   & $1089$ & $6643$ & $4.95\cdot10^4$ & $1.71\cdot10^6$ \\ 
  && \textbf{Ours}    & \textbf{$\mathbf{108.4}$}& \textbf{$\mathbf{2972}$} & \textbf{$\mathbf{3888}$} & $\mathbf{1.53\cdot10^4}$ & $\mathbf{3.47\cdot10^4}$ & $\mathbf{15.91}$   & $\mathbf{1.8\cdot10^5}$& $\mathbf{1.12\cdot10^6}$ & $\mathbf{4.09\cdot10^7}$ & $\mathbf{2.39\cdot10^8}$ \\ \cline{2-13}

  & \multirow{6}{*}{FactorJoin} 
  & Clean& $3.015$& $6.82$& $39.11$ &  $1.36\cdot10^4$& $1.98\cdot10^5$  
  & $2.018$   & $40.61$ & $78.91$ & $302.7$ & $760.1$ \\  
  && Random& $3.176$& $5.481$& $45.82$& $1.63\cdot10^4$& $8.01\cdot10^5$ 
  & $2.072$   & $45.98$ & $64.56$ & $312.98$ & $774.3$ \\  
  && GA   & $4.088$ & $156.7$ & $4087$ & $1.19\cdot10^5$ & $1.54\cdot10^6$  
  & $5.03$  & $115.8$ &$ 296.2 $& $921.8$ & $1.13\cdot10^4$ \\  
  && SA   & $3.999$ & $889.6$ & $1.76\cdot10^4$& $8.76\cdot10^5$ & $3.68\cdot10^8$ & $6.18$   & $784.6$ & $1.67\cdot10^4$ & $2.22\cdot10^7$& $1.41\cdot10^9$  \\  
  && PACE& $3.015$& $6.82$& $39.11$ &  $1.36\cdot10^4$& $1.98\cdot10^5$  
  & $2.018$   & $40.61$ & $78.91$ & $302.7$ & $760.1$ \\ 
  && \textbf{Ours}    & $\mathbf{6.706}$& $\mathbf{1.56\cdot10^5}$ & $\mathbf{2.33\cdot10^6}$& $\mathbf{4.523\cdot10^7}$& $\mathbf{1.04\cdot10^9}$ & $\mathbf{6.486}$  & $\mathbf{2099}$& $\mathbf{8.05\cdot10^4}$ & $\mathbf{1.72\cdot10^8}$ & $\mathbf{4.52\cdot10^9}$ \\ \hline

\multirow{12}{*}{Hybrid}& \multirow{6}{*}{ALECE}      & Clean& $1.374$& $2.611$ & $3.702$ & $16.03$ & $649.4$  & $1.532$   & $3.007$ & $4.321$ & $19.55$ & $48.89$ \\ 
  && Random& $1.566$ & $2.741$ & $3.499$ & $9.782$ & $439.7$  & $1.648$   & $3.563$& $4.583$ & $27.04$& $68.4$ \\  
  && GA   & $13.86$ & $159.2$ & $212.9$ & $5547$ & $1.28\cdot10^5$ & $4.877$   & $76.31$ & $186.9$ & $417.5$ & $2027$ \\  
  && SA   & $27.01$ & $696.1$ & $2723$ & $5.06\cdot 10^4$ & $1.49\cdot10^5$ & $4.32$   & $724.5$ & $1405$ & $3458$ & $4631$ \\  
  && PACE & $7.229$ & $134.4$ & $265.6$& $4284$ & $4.62\cdot10^5$ & $4.08$  & $207.5$ & $544.8$ & $5637$ & $3.03\cdot 10^4$ \\  
  && \textbf{Ours}    & $\mathbf{29.33}$ & $\mathbf{9323}$ & $\mathbf{2.73\cdot10^4}$& $\mathbf{2.45\cdot10^5}$ & $\mathbf{2.02\cdot10^6}$  & $\mathbf{6.442}$   & $\mathbf{3.23\cdot 10^4}$ & $\mathbf{2.06\cdot10^5}$& $\mathbf{5.88\cdot10^6}$ & $\mathbf{1.64\cdot10^7}$ \\ \cline{2-13}

  & \multirow{6}{*}{RobustMSCN} & Clean& $1.811$& $4.415$& $23.67$& $355.5$ & $1102$ & $2.288$   & $5.196$ & $7.43$& $28.62$ & $4470$ \\  
  && Random& $2.145$& $5.409$& $23.81$& $381.5$& $2547$ & $2.723$    & $5.903$& $7.743$ & $23.55$& $4656$ \\  
  && GA   & $15.30$ & $109.1$ & $173.8$ & $627.8$ & $1.83\cdot10^4$ & $3.820$   & $47.62$ & $81.41$& $274.9$& $7.38\cdot10^4$ \\  
  && SA   & $12.69$& $186.3$& $805.5$& $5705$& $2.16\cdot10^4$ & $3.92$   & $83.2$& $270.5$& $1225$& $2.85\cdot 10^4$ \\  
  && PACE & $5.019$& $77.26$ & $278.7$& $2352$ & $7155$ & $4.84$   & $98.18$ & $426.7$& $4009$& $1.48 \cdot 10^4$ \\  
  && \textbf{Ours}    & $\mathbf{15.87}$& $\mathbf{814.7}$& $\mathbf{2185}$& $\mathbf{1.52\cdot10^4}$& $\mathbf{2.64\cdot10^5}$ & $\mathbf{5.19}$   & $\mathbf{2388}$& $\mathbf{8762}$& $\mathbf{1.32\cdot10^5}$ & $ \mathbf{7.06\cdot10^{5}} $  \\ \hline
  
\multicolumn{1}{l|}{\multirow{5}{*}{}}
& \multirow{5}{*}{Oracle}     & Clean& 1& 1 & 1& 1 & 1& 1& 1& 1&1& 1 \\  
\multicolumn{1}{l|}{}  & 
& Random  & 1.255 & 1.294 & 1.663 & 1.766 &  3.338 
& 1.212    & 1.298& 1.319& 1.515& 2.328 \\  
\multicolumn{1}{l|}{}  & & GA   & 83.39& 230.3 & 2005 & $5.15 \cdot 10^5$ & $1.674 \cdot 10^6$
& 25.97& 258.8 & 1566 & $8741$ & $1.44 \cdot 10^4$ \\  
\multicolumn{1}{l|}{}  & & SA   & 301.1 & $2.72\cdot 10^5$  & $5.12\cdot 10^5$& $1.54\cdot 10^7$ & $3.23\cdot 10^7$ 
& 35.98  & 2133 & 4835 & $1.43 \cdot 10^4$ & $1.56 \cdot 10^6$ \\  
\multicolumn{1}{l|}{}  & & \textbf{Ours}      & $\mathbf{1.49\cdot 10^5}$        & $\mathbf{1.23\cdot 10^8}$& $\mathbf{5.03\cdot 10^8}$& $\mathbf{3.89\cdot 10^9}$& $\mathbf{9.53\cdot 10^9}$
& $\mathbf{1.66 \cdot 10^4}$   & $\mathbf{1.02 \cdot 10^8}$& $\mathbf{2.50 \cdot 10^8}$& $\mathbf{2.21 \cdot 10^9}$  & $\mathbf{1.78 \cdot 10^{10}}$ \\ \hline

\end{tabular}

}
\end{table*}

%



\textbf{Percentile Decline of the Estimator's Qerror:} The distribution of estimator errors can be analyzed through Qerror percentiles, where the 50-th percentile provides an alternative measure of overall estimation quality~\cite{yang2019deep}, and higher percentiles directly impact database query optimization performance~\cite{PACE,PACE_Percentile_Proof}. Table~\ref{SuperBigTab} presents the percentile Qerror values for each CE model before and after adversarial attacks across multiple datasets. Our approach achieves a 50-th percentile Qerror reduction of $8.28\times$, $2.43\times$, $6.47\times$, and $47.1\times$ compared to PACE, SA, GA, and Random, respectively. Furthermore, on Qerror percentiles exceeding 90-th, our method exhibits a 3–4 orders of magnitude improvement over all baselines. Table~\ref{SuperBigTab} also includes the attack performance against the training database oracle $E_O$. Our method demonstrates the strongest misleading effect, theoretically increasing Qerror by 5 orders of magnitude for over 50\% of queries. Notably, high-accuracy estimators are particularly vulnerable to DACA attacks. For instance, Neurocard—the top-performing model on IMDB—experiences a 2 orders of magnitude Qerror increase for over 50\% of queries when attacked by our method. This vulnerability arises because such models closely mimic the oracle within training data, making them more susceptible to DACA perturbations. These results validate the theoretical transformations discussed in \S~\ref{ATK_ORC}.\looseness=-1

\textbf{Decline of the E2E-Time:} To systematically evaluate how different attack methods impact database query optimization pipelines, we measured the end-to-end execution time of various cardinality estimators under different attack strategies. Figures~\ref{Fig.IMDB_PerQueryE2E} and~\ref{Fig.STATS_PerQueryE2E} present the performance degradation analysis for STATS-CEB and IMDB-JOB queries, respectively. Our method consistently demonstrates superior effectiveness in compromising all tested estimators, misleading optimizers to produce the worst execution plans. Specifically, our approach achieves execution time increments of $24.2\times$, $2.21\times$, $17.5\times$, and $202\times$ over PACE, SA, GA, and Random on IMDB-JOB, and $11.6\times$, $2.34\times$, $10.3\times$, and $41.8\times$ on STATS-CEB. Analysis reveals varying susceptibility among estimators to data-centric attacks regarding E2E-Time, ranked as: LWNN > MSCN > NeuroCard > ALECE > FactorJoin > RobustMSCN. The hybrid RobustMSCN shows relative resilience (only $1.58\times$ slowdown) due to its robust feature utilization, though all estimators remain vulnerable. These results conclusively demonstrate our attack's effectiveness in corrupting estimator knowledge, forcing optimizers to select suboptimal plans with significant performance penalties.
\looseness=-1







\begin{yellowbox}
\textbf{Takeaways}: Our DACA methodology  effectively compromises the performance of all kinds of learned estimators. It successfully corrupted the knowledge acquired by different learned models, whether they are data-driven, query-driven, or hybrid models. Although our DACA methodology was originally designed to reduce the average Qerror. We are surprised to find out that it can also significantly diminish the model's worst-case prediction accuracy, misguiding the optimizer to generate suboptimal physical execution plans and resulting in multiple-fold increases in end-to-end latency. Furthermore, we observe that the data-level robustness features employed by the latest hybrid models can mitigate some of the attack's effects; however, the effectiveness of this mitigation remains to be enhanced.\looseness=-1
\end{yellowbox}

\subsection{Scalability Study}
\label{Sec.ScaEva}

\begin{figure}[htbp]
 \subfigure[ Attack Effectiveness on different tables. ]{
		\label{Fig.TabEffect}
		\includegraphics[width=0.7 \columnwidth]{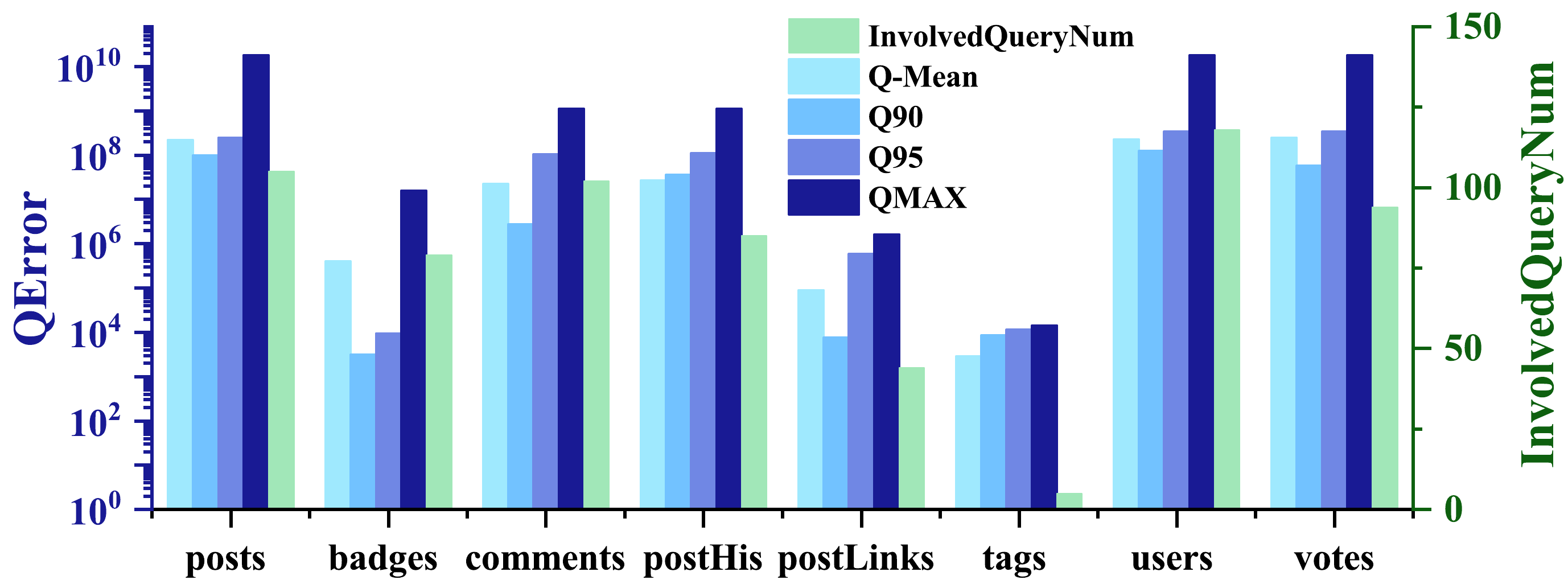}
	}  
    \vspace{-1.4em}    
	\subfigure[ Attack Efficiency on different tables. ]{
		\label{Fig.Efficiency}
		\includegraphics[width=0.7 \columnwidth]{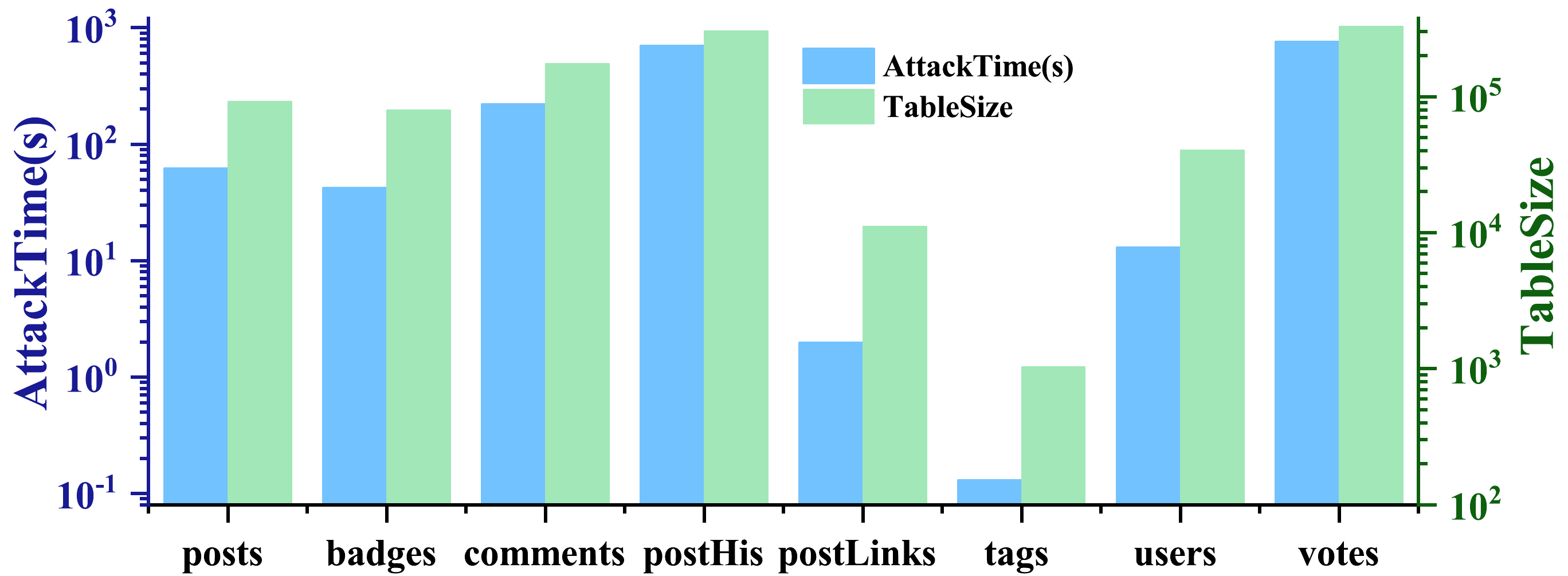}
	}
	
	\caption{ Scalability on different tables.}
	\vspace{-1.4em}    
\end{figure}


\textbf{Scalability on Attacked Tables:} To validate the theoretical attack effectiveness on different tables, in Figure~\ref{Fig.TabEffect}  and Figure~\ref{Fig.Efficiency}, we conducted experiments by switching the attacked tables within the STATS-CEB database. We selected the oracle \(E_{O}\) as the victim. The attack budget was set at 20\% of the rows in the targeted table. Our findings indicate that, despite the varying number of queries associated with different tables in the STATS-CEB workload, our method can significantly compromise \(E_{O}\) for any table, resulting in a maximum Qerror ranging from \(10^4\) to \(10^{10}\). Additionally, we observed that the more queries a table encompasses, the greater the impact of the attack. For instance, when the number of queries increased from 10 (in \texttt{tags}) to 76 (in the \texttt{votes}), the maximum Qerror escalated from \(10^4\) to \(10^{10}\).\looseness=-1


\vspace{-1.0em} 
\begin{figure}[htbp]
	\subfigure[ Attack budget vs Qerror. ]{
		\label{Fig.BudgetScale}
		\includegraphics[width=0.4\columnwidth]{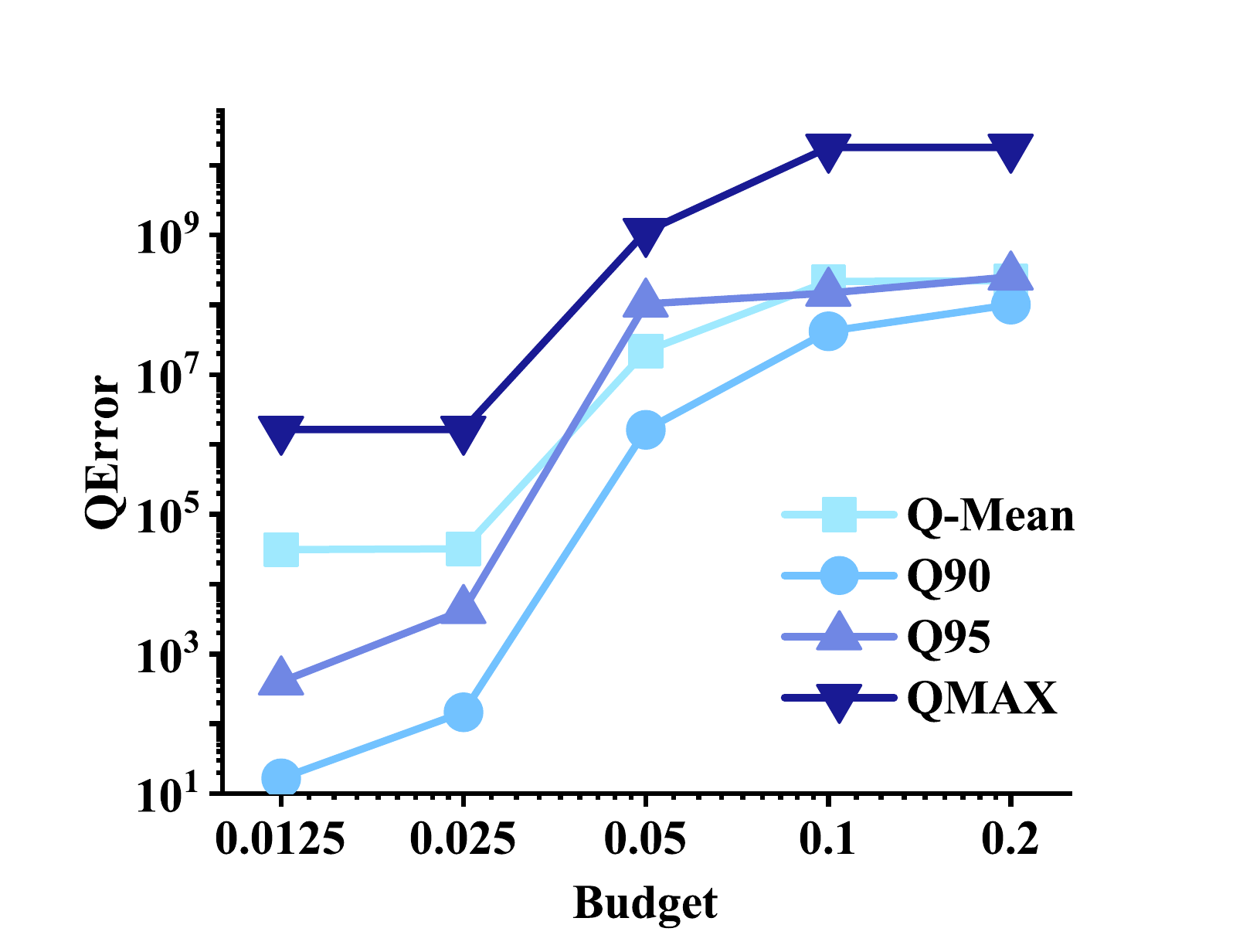}
	}
	\subfigure[ Scalability and E2E-Time. ]{
		\label{Fig.TimeScale}
		\includegraphics[width=0.4\columnwidth]{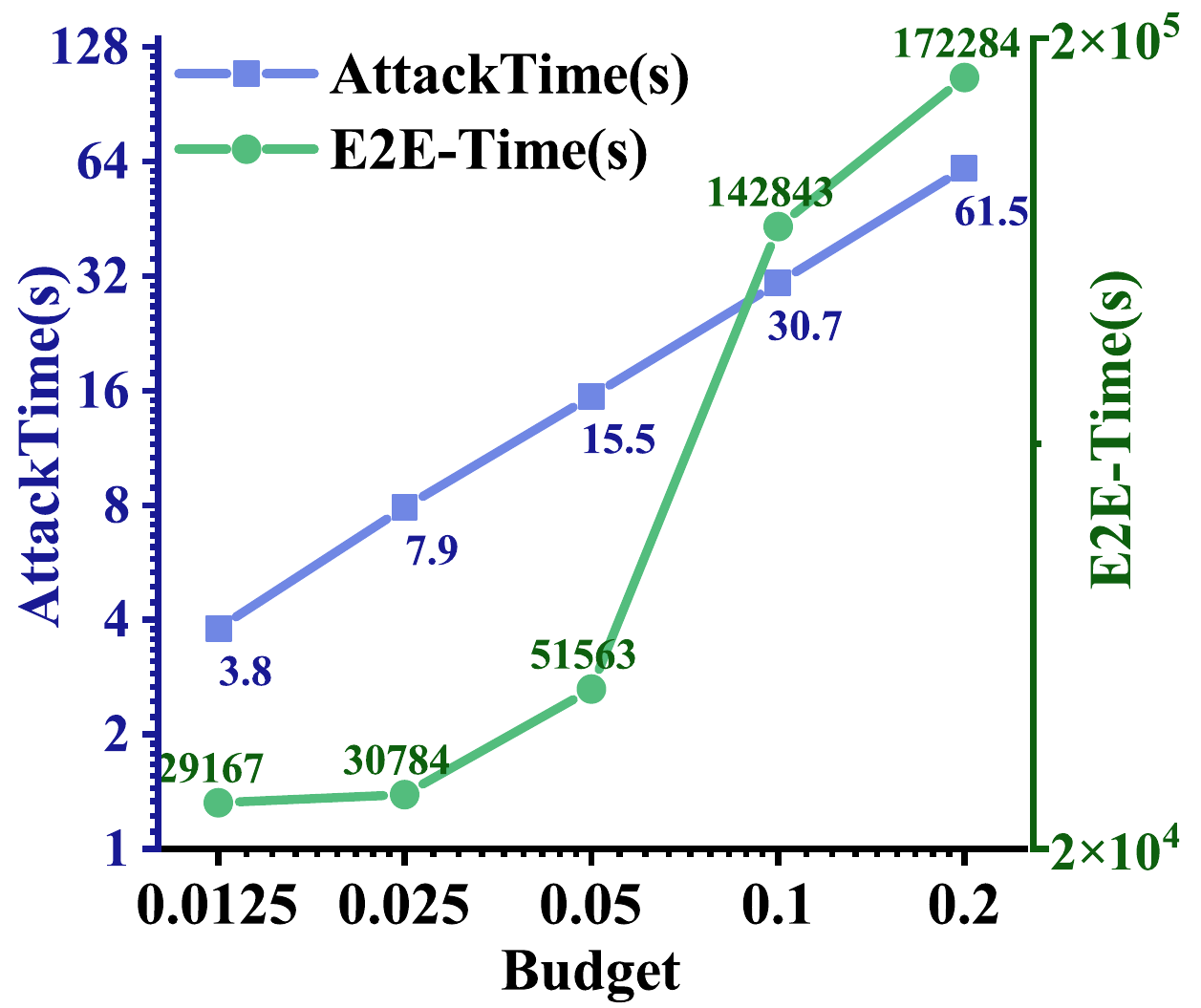  }
	}
    \vspace{-1.0em} 
	\caption{ Scalability on different budget.}
	\label{Exp03.bud}
	\vspace{-1.0em}    
\end{figure}


\textbf{Scalability on Attacked Budget}: To test the theoretical attack effectiveness on different attack budgets, we conducted experiments by switching the attack budget in the \texttt{posts} table of the STATS-CEB database from $1.25\%$ to $20\%$. We choose the $E_{O}$ as the victim estimator. Figure~\ref{Fig.BudgetScale} and ~\ref{Fig.TimeScale} illustrates the effectiveness and duration of our attack. It can be observed that the attack time is proportional to the attack budget, demonstrating good scalability and performance. Additionally, a larger attack budget, while requiring more time, also yields better attack effectiveness. For instance, increasing the budget from 0.05 to 0.1 results in an end-to-end time increase by $2.7\times$.\looseness=-1


\begin{yellowbox}
\textbf{Takeaways}: The overhead of our proposed DACA attack strategy is linear to the attack budget and correlated with table size and query number. Moreover, the more queries that involve the attacked table, the larger the attack budget, and consequently, the greater the benefit derived from the attack.    
\end{yellowbox}

\subsection{Consequence and Possible Defense}
\label{Sec.Consequence}
\begin{figure}[htbp]

 \subfigure[  Distribution on: IMDB-JOB.  ]{
		\label{Fig.IMDB_Case}
		\includegraphics[width=0.4\columnwidth]{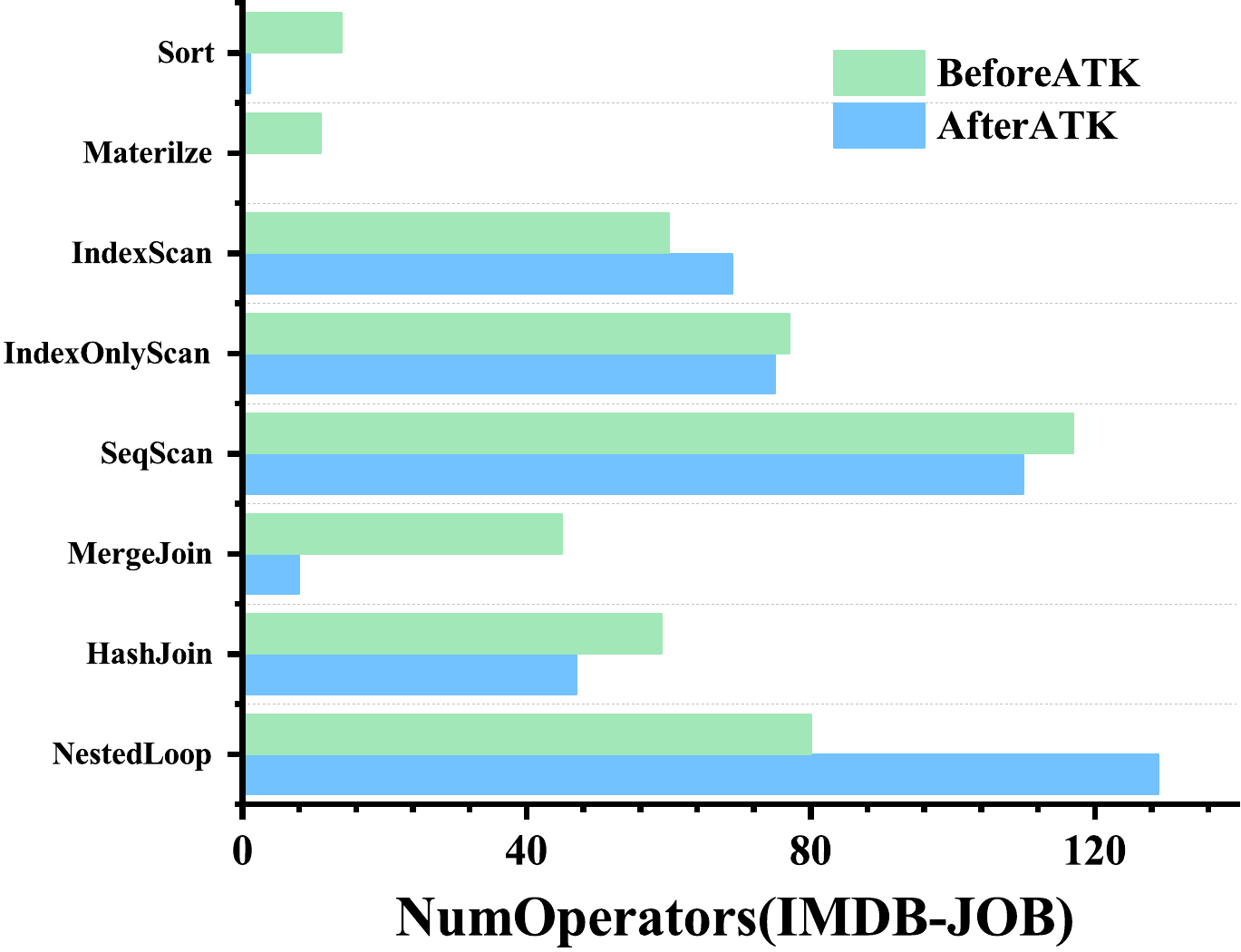}
	}  
	\subfigure[   Distribution on: STATS-CEB.  ]{
		\label{Fig.STATS_Case}
		\includegraphics[width=0.4\columnwidth]{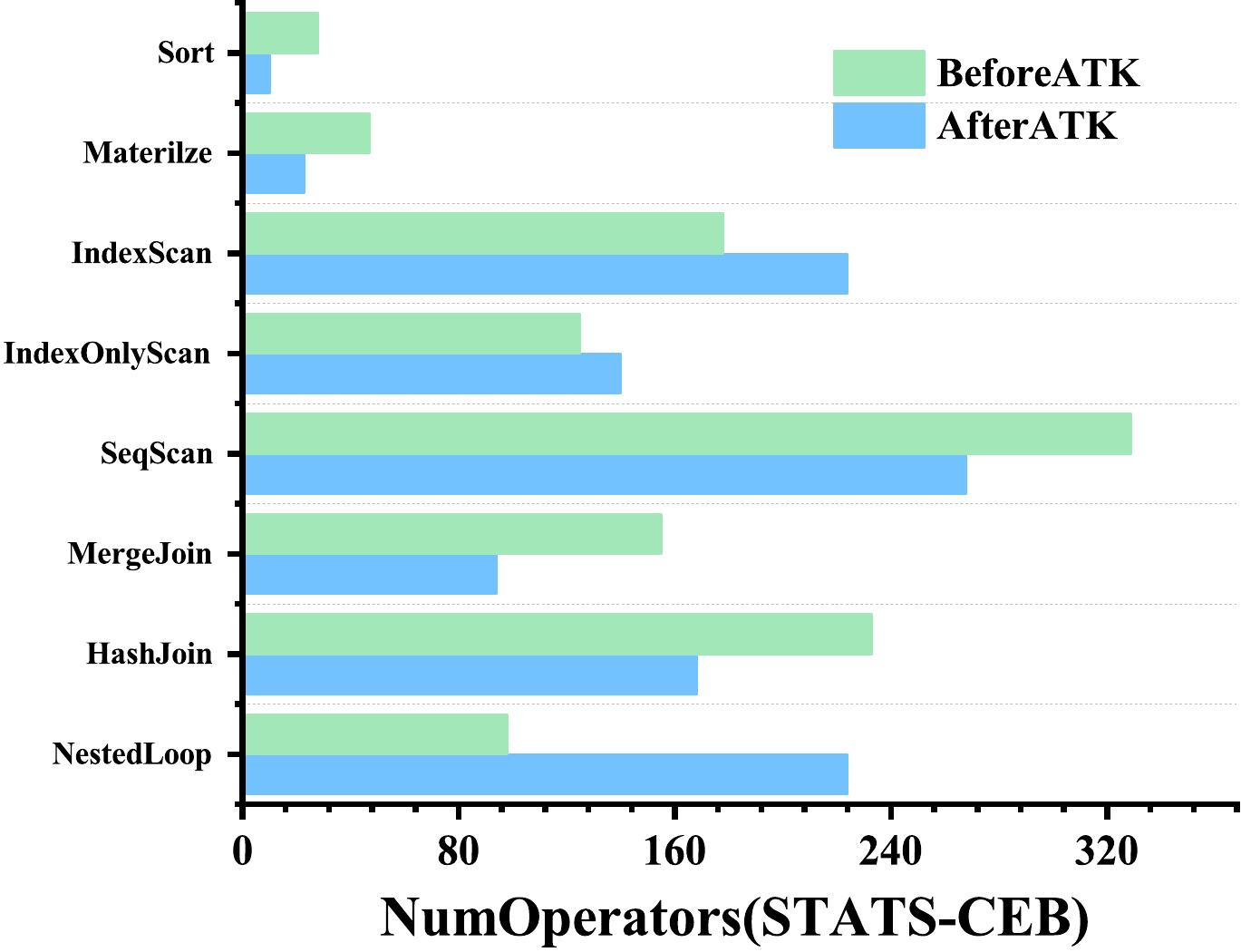}
	}
    \vspace{-1.0em}  
	\caption{ Operator Distribution. }
	\label{Exp.OperDis}
	\vspace{-1.0em}    
\end{figure}



 
In this section, we studied the consequence of DACA and possible defense strategies against DACA. For consequence study, we have selected LWNN as the target model. We conducted tests on IMDB-JOB and STATS-CEB, and we analyzed the distribution of physical plans. We visualized the operators' distribution in Figure~\ref{Exp.OperDis}. 

\begin{figure}[htbp]

\subfigure[ Before Attack. ]{
		\label{Fig.BeforATK}
		\includegraphics[width=0.3\columnwidth]{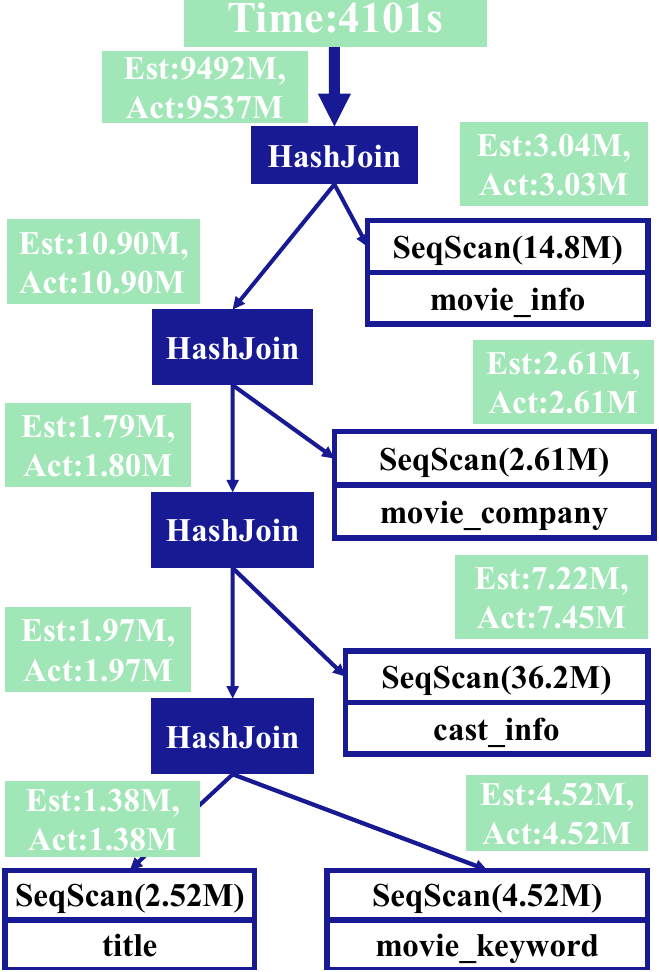}
	}
 \subfigure[ After Attack. ]{
				\label{Fig.AfterATK}
		\includegraphics[width=0.37\columnwidth]{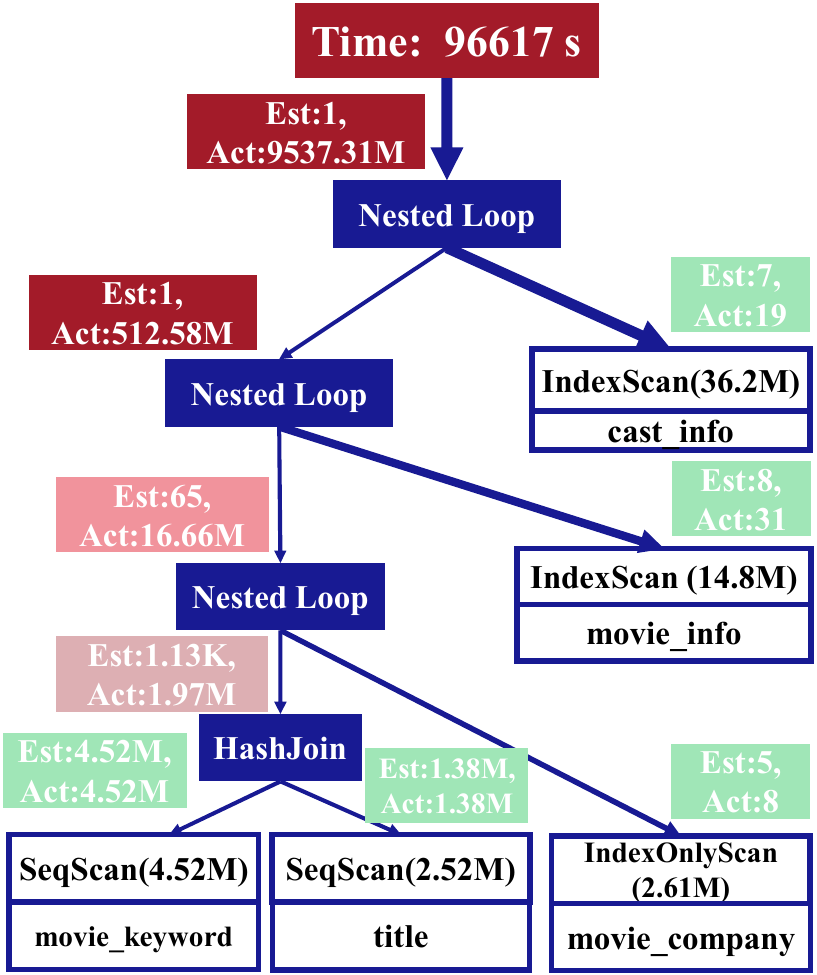}
	}  
	\vspace{-1.0em}
	\caption{ Physical plan of IMDB-JOB Q60. }
\end{figure}

\textbf{Join Pattern:} Our analysis reveals that post-attack execution plans exhibited significant shifts in operator usage compared to pre-attack plans. Specifically, Nested Loop and Index Scan operators showed marked performance improvements, with Nested Loop performance increasing by 128\% and 61\% in the IMDB-JOB and STATS-CEB databases, respectively, and Index Scan performance rising by 12\% and 15\% in these databases. Conversely, Hash Join and Merge Join operators experienced substantial reductions, decreasing by 27\% and 20\% (Hash Join) and 18\% and 32\% (Merge Join) in the respective databases. These findings indicate that erroneous cardinality estimates post-attack severely misled the query optimizer into favoring Nested Loop Joins or Index-Based Nested Loop Joins—operators better suited for small-scale queries—over more efficient Hash Joins or Merge Joins for complex analytical workloads. A representative example is the degradation of IMDB-JOB Q60, as shown in Figure~\ref{Fig.BeforATK} and Figure~\ref{Fig.AfterATK}. Here, the misled cardinality estimator opts for Index Nested Loop joins across all three top nodes of the query plan tree instead of Hash Joins. This operator misselection increases computational complexity, thereby extending end-to-end query processing time.\looseness=-1

\textbf{Cache Pattern:} Additionally, we observed that, aside from the common joint operators, the proportion of materialized operators also significantly decreased. In the IMDB-JOB database, the use of materialize operators was reduced by 90\%, and in the STATS-CEB database, their proportion decreased by 51\%. This suggests that the compromised cardinality estimator misleads the optimizer to reduce access to materialized caches, thereby increasing the rate of redundant computations and decreasing cache hit rates.

\begin{figure}[htbp]
    \vspace{-1.0em}    
    \subfigure[  Ensemble on: IMDB-JOB.  ]{
		\label{Fig.IMDB_Ensemble}
		\includegraphics[width=0.4\columnwidth]{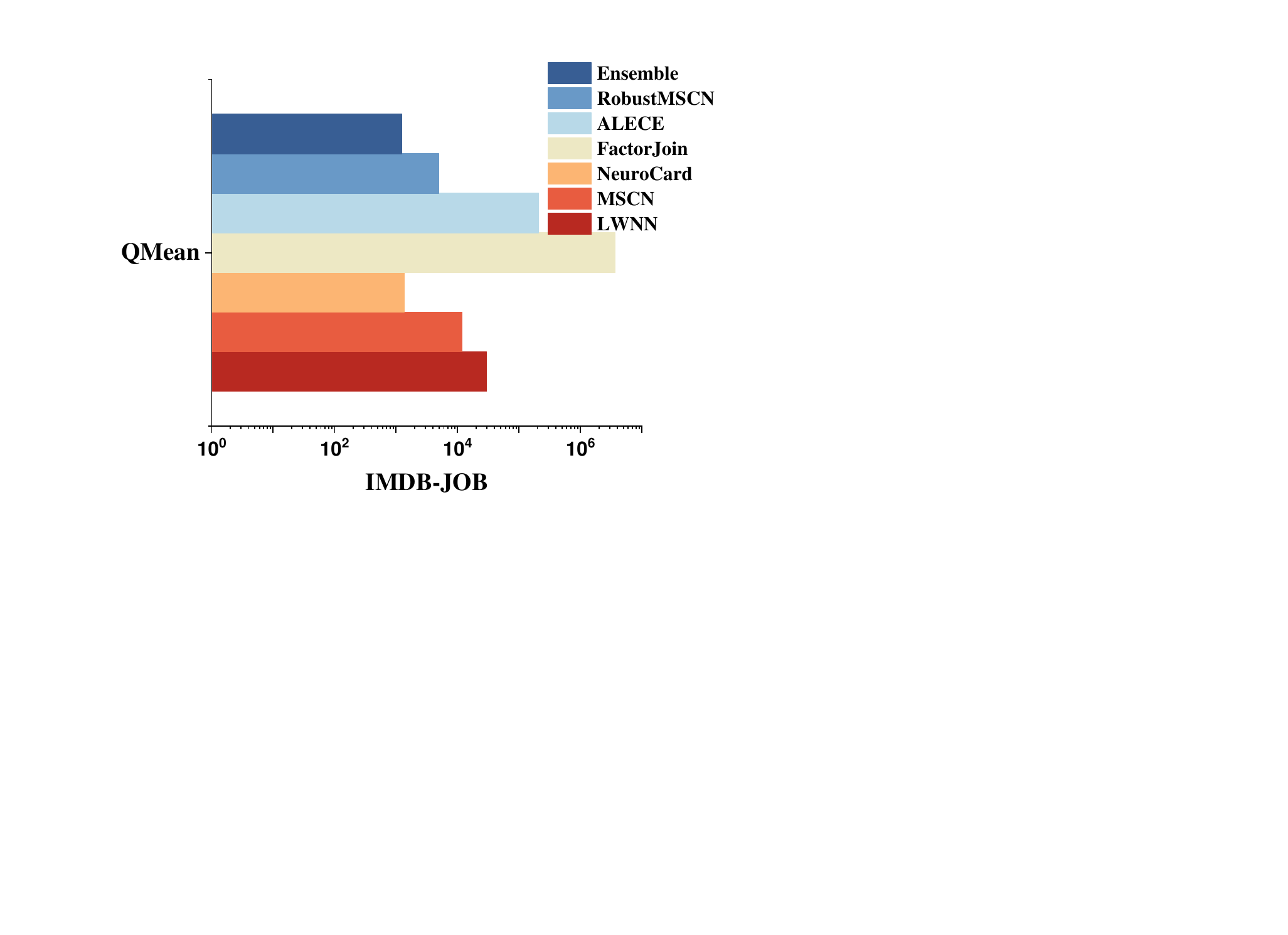}
	}  
	\subfigure[    Ensemble on: STATS-CEB.  ]{
		\label{Fig.STATS_Ensemble}
		\includegraphics[width=0.4\columnwidth]{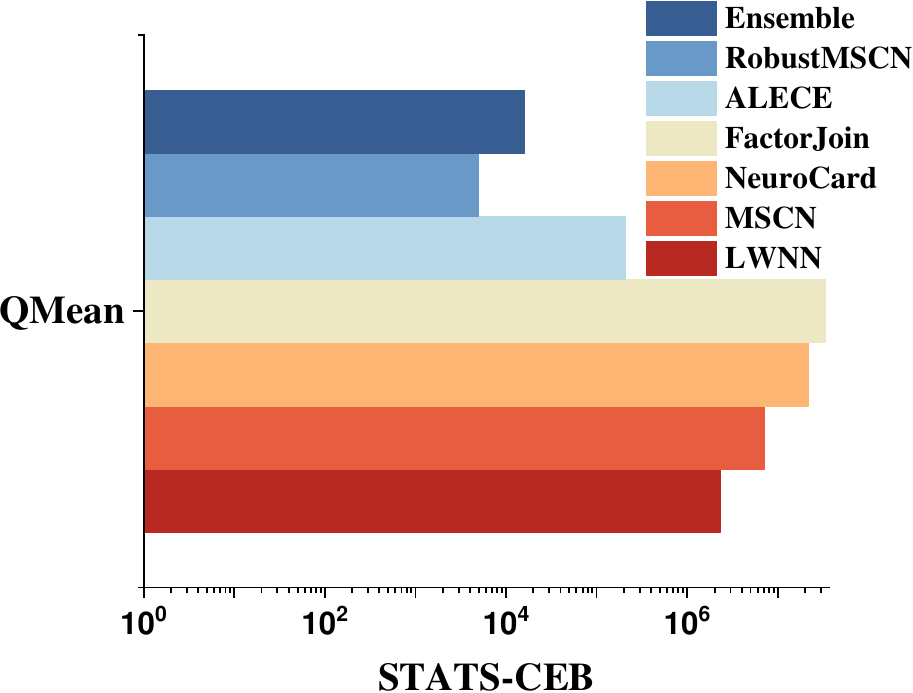}
	}
        \vspace{-1.0em}    
	\caption{ Countermeasure Study: Ensembling.  }
	\label{Exp.Ensemble}
	\vspace{-1.0em}    
\end{figure}

%

\textbf{Countermeasure:} In Figure~\ref{Fig.IMDB_Ensemble},  we discuss one possible defense against DACA. Due to space constraints, we present one possible countermeasure in the main text. Additional defenses and their limitations are discussed in our appendix~\cite{DACA_Appendix}. Inspired by existing researches~\cite{xu2025adanndv,RobustMSCN}, we introduce redundancy into the deployed   estimators. Specifically, we train an additional MLP to implement an integration scoring function, which performs a weighted summation of the estimation results from the six estimators used in our experiments. This approach yields a robust integrated result. We observe that the integrated estimator indeed achieves the most robust performance, attaining optimal or near-optimal results in two post-attack scenarios. However, we note that the integrated estimation model does not significantly outperform all the individual models being integrated. This indicates that while simple integration of redundant models can mitigate the worst-case performance of a single estimator to some extent, achieving a qualitative improvement in worst-case performance calls for further study.\looseness=-1

\begin{yellowbox}
\textbf{Takeaways}: After being subjected to our data-centric algorithm complexity attack, the learned estimator supplies inaccurate estimates to the database optimizer, which has two major consequences: (1) \textbf{Surge in computational complexity:} The mislead optimizer tends to select the suboptimal Nested Loop operators for hard analytical queries, thereby increasing the theoretical time complexity and prolonging execution times. (2) \textbf{Decline in memory cache hit rates:} The misguided optimizer tends to expand computations from scratch rather than reuse existing materialized computational results,  which subsequently lowers the system's cache hit rate and diminishes overall system efficiency.  Meanwhile,  our further countermeasure study demonstrates that combining multiple estimators via learned weights achieves the certain defense against DACA, although qualitative improvements may necessitate more sophisticated future research.
\end{yellowbox}


\section{Related Work}


\textbf{Learned Cardinality Estimators}:  Modern cardinality estimators can be  categorized into three paradigms: query-driven~\cite{LWNN,MSCN_Kipf2018LearnedCE,UAE_Q}, data-driven\cite{yang2019deep,hilprecht2019deepdb,CardIndex,nec,FactorJoin,CardIndex} and hybrid~\cite{PRICE,ALECE,RobustMSCN}. \textbf{(1) Query-driven estimators} employ neural networks ~\cite{LWNN,MSCN_Kipf2018LearnedCE,UAE_Q} to learn mappings from query representations to true cardinalities. Although query-driven methods can achieve quick and lightweight estimations, it is confirmed by many studies that query-driven estimators suffer from changes in testing query distribution, which may be caused by data drifts or malicious poisoning within query workloads.~\cite{AreWeReady4CE,yang2019deep,ALECE,wang2021face}. \textbf{(2)Data-driven estimators} learn joint data distributions and sample from deep models~\cite{yang2019deep,nec,ASM} to estimate cardinalities.  They are robust against workload drifts~\cite{yang2019deep,AreWeReady4CE}. \textbf{(3) Hybrid estimators}~\cite{RobustMSCN,ALECE,wu2021unified,MSCN_Kipf2018LearnedCE,RobustMSCN,Hilprecht2022a} use statistical information from the data and query representations for unified learning. Commonly used data features include histograms~\cite{ALECE,PRICE}, sample collections~\cite{MSCN_Kipf2018LearnedCE,RobustMSCN}, and estimations from PostgreSQL~\cite{PRICE,RobustMSCN,Hilprecht2022a}.

\textbf{Attacks on Learned Database Components:} Nowadays, attacks on learned database components have garnered widespread attention~\cite{PoisonLI,ACA_Learnedindex,ATTACK_IndexAdvisor,PACE}. Kornaropoulos et al.~\cite{PoisonLI,ACA_Learnedindex} introduced poisoning attacks and algorithm complexity attacks targeting learned indexes by constructing malicious inputs that cause learned indexes to fail. Additionally, Zheng et al.~\cite{ATTACK_IndexAdvisor} proposed poisoning attacks aimed at learning-based index advisors. The work most closely related to ours is PACE proposed by Zhang et al.~\cite{PACE}, which affected query-driven estimators by adding noise to their training workloads. However, PACE has a clear limitation: it can only target query-driven models. In contrast, the attack method proposed in this paper can influence nearly all types of cardinality estimators.\looseness=-1


\textbf{Data Poisoning Attacks:} In machine learning, poisoning attacks are typically analyzed under the white-box assumption, where the attacker has full knowledge of the internal model~\cite{ACA8,ACA29,ACA65,PACER0,PACER1,PACERX,PACERY}. For example: (1). Under the assumption of linear regression, Jagielski et al.~\cite{ACA29} proposed an optimization framework for poisoning attacks targeting linear regression models. (2). For support vector machines (SVMs)~\cite{ACA8,PACERX,PACERY}, maliciously crafted training data can manipulate the decision boundary, thereby increasing the error rate. (3). In the context of neural networks, Yang et al.~\cite{ACA65} introduced a gradient-based technique to generate poisoning inputs. In contrast, our work focuses on a more realistic black-box scenario, where the attacker lacks knowledge of the internal model details. Specifically, the attacker is unaware whether the internal model is based on neural networks~\cite{nec,ALECE}, decision trees~\cite{hilprecht2019deepdb,LWNN,Flat}, Bayesian networks~\cite{FactorJoin,BayesCard}, or other emerging models~\cite{wang2021face,PRICE}.\looseness=-1

\vspace{-1.0em}


\section{Conclusion}
This study reveals critical vulnerabilities of learned cardinality estimators to minimal data-level drifts. Through a black-box data-centric algorithmic complexity attack, we demonstrate that even slight training data modifications can severely degrade estimator accuracy. We prove the NP-Hardness of finding optimal attack strategies and develop an efficient $(1-\kappa)$-approximation algorithm with polynomial-time complexity. Experiments on STATS-CEB and IMDB-JOB benchmarks show our attack's effectiveness: modifying merely 0.8\% of database tuples causes a 1000$\times$ increase in 90-th percentile Qerror and up to 20$\times$ longer processing times. These findings expose the fragility of current estimators in real-world scenarios. We also propose practical countermeasures against such black-box attacks, providing valuable insights for developing robust learned optimizers. Future work could focus on creating more resilient estimation techniques that can withstand data-level disruptions while maintaining reliable performance.\looseness=-1


\bibliographystyle{ACM-Reference-Format}
\bibliography{sample}

\appendix
\clearpage 

Due to space limitations, we present some proof details in this appendix.

\section{Proof of Theorem3.2}
We present the proof of Theorem 3.2 here:
    %
\begin{proof}
We aim to prove that when \( x > K \times M \), the benefit in Qerror resulting from inserting \( I \) tuples (\(2 \leq I \leq K\)) into the database constructed from Theorem 3.1 is smaller than the benefit obtained by not deleting any two tuples corresponding to any \( q_x \). Consequently, the plan of inserting \( I \) tuples can be disregarded and replaced with \( I \) delete operations. And we can therefore reuse the PTIME reduction in theorem\ref{thm:optimal-data-level-attack-np-hard}, construct a polynomial-time reduction from the DKS problem to the DACA problem when simultaneously considering insertions and deletions. Therefore the DACA problem considering both insertions and deletions remains NP-Hard.

For a fixed query \( q_x \), deleting its two tuples in \( R_x \) yields a benefit \( g_{\text{Del}} = 2x + 1 \). Given that \( x > K \times M \), we have
\[
g_{\text{Del}} > 2 \times K \times M + 1.
\]

Assume that all \( M \) queries share one common test tuple in \( R_x \). Therefore, repeatedly inserting into this special tuple would be the optimal solution for inserting \( I \) tuples, which provides an upper bound on the Qerror when inserting \( I \) tuples. That is,
\[
g_{\text{Ins}} \leq M \times \frac{I + 2}{2}.
\]

Given that \( g_{\text{Del}} > 2 \times K \times M + 1 \) and \( K \geq I \), we have
\[
g_{\text{Del}} > 2 \times K \times M > 2 \times I \times M = \frac{I M}{2} + \frac{3 I M}{2}.
\]
Since \( I \geq 2 \), we have \( \frac{3 I M}{2} > M \), and thus
\[
2 \times I \times M = \frac{I M}{2} + \frac{3 I M}{2} > \frac{I M}{2} + M \geq g_{\text{Ins}}.
\]
In conclusion, we deduce that the benefit \( g_{\text{Del}} \) obtained by deleting any two tuples corresponding to query \( q_x \) in \( R_x \) is greater than the benefit \( g_{\text{Ins}} \) of inserting \( I \) tuples (\(2 \leq I \leq K\)).
\end{proof}

\section{Proof of the derivation in Theorem4.1}

Given that 
$ \mathbb{Q}_j(X) = \frac{C_D(\mathbf{q}_j) + 1}{C_D(\mathbf{q}_j) + 1 + \sum_{t_i\in X} t_i.\beta\times w_{ij}}$
and 

$w'_{ij} = \frac{w_{ij}}{1 + C_D(\mathbf{q}_j )} ,\quad S_j(X) = \sum_{t_i\in X} t_i.\beta w'_{ij}$
    
We prove that  the expression
    \[
    \mathbb{Q}_j(A \cup B) + \mathbb{Q}_j(A \cap B) - \mathbb{Q}_j(A) - \mathbb{Q}_j(B)
    \]

can be reorganized in the following form:
   
\[
\quad \frac{(2+S_j(A)+S_j(B))(S_j({A \setminus B}))(S_j({B \setminus A}))}{(1 + S_j({A \cup B}))(1 + S_j({A \cap B}))(1 + S_j(A))(1 + S_j(B))}
\]
\begin{proof}
    
    
    Based on the weights \( w'_{ij} \) and the function \( S_j(X) \), we define the function \( \mathbb{Q}_j(X) \) as follows:

\begin{equation*}
\mathbb{Q}_j(X) = \frac{C_D(\mathbf{q}_j) + 1}{C_D(\mathbf{q}_j) + 1 + \sum_{t_i\in X} t_i.\beta\times w_{ij}} = \frac{1}{1 + S_j(X)}.
\end{equation*}

Therefore, for any sets \( A \) and \( B \), the following relationship holds:

\begin{align*}
    & \mathbb{Q}_j(A \cup B) + \mathbb{Q}_j(A \cap B) - \mathbb{Q}_j(A) - \mathbb{Q}_j(B)  \\
    =& \frac{1}{1 + S_j(A \cup B)} + \frac{1}{1 + S_j(A \cap B)} - \frac{1}{1 + S_j(A)} - \frac{1}{1 + S_j(B)} 
\end{align*}
We examine the first two terms of the above summation:

\begin{equation*}
\frac{1}{1 + S_j(A \cup B)} + \frac{1}{1 + S_j(A \cap B)} 
\end{equation*}

This can be combined as follows:

\begin{align*}
&\frac{1}{1 + S_j(A \cup B)} + \frac{1}{1 + S_j(A \cap B)} \\
&= \frac{(2 + S_j(A \cup B) + S_j(A \cap B)) \times (1 + S_j(A)) \times (1 + S_j(B))}{(1 + S_j(A \cup B))(1 + S_j(A \cap B))(1 + S_j(A))(1 + S_j(B))} \\
&= \frac{2 + S_j(A \cup B) + S_j(A \cap B)}{(1 + S_j(A \cup B))(1 + S_j(A \cap B))} \times \frac{(1 + S_j(A))(1 + S_j(B))}{(1 + S_j(A))(1 + S_j(B))}
\end{align*}

Similarly, the last two terms of the summation are:

\begin{equation*}
\frac{1}{1 + S_j(A)} + \frac{1}{1 + S_j(B)}
\end{equation*}

These terms can be combined as:

\begin{align*}
&\frac{1}{1 + S_j(A)} + \frac{1}{1 + S_j(B)} \\
&= \frac{(2 + S_j(A) + S_j(B)) \times (1 + S_j(A \cup B)) \times (1 + S_j(A \cap B))}{(1 + S_j(A))(1 + S_j(B))(1 + S_j(A \cup B))(1 + S_j(A \cap B))} \\
&= \frac{2 + S_j(A) + S_j(B)}{(1 + S_j(A))(1 + S_j(B))} \times \frac{(1 + S_j(A \cup B))(1 + S_j(A \cap B))}{(1 + S_j(A \cup B))(1 + S_j(A \cap B))}
\end{align*}

Notably, we observe that:

\begin{equation*}
2 + S_j(A) + S_j(B) = 2 + S_j(A \setminus B) + 2S_j(A \cap B) + S_j(B \setminus A) = 2 + S_j(A \cup B) + S_j(A \cap B)
\end{equation*}

Therefore, by combining the above results, we obtain:

\begin{align*}
&\frac{1}{1 + S_j(A \cup B)} + \frac{1}{1 + S_j(A \cap B)} - \frac{1}{1 + S_j(A)} - \frac{1}{1 + S_j(B)} = \\
&{\footnotesize\frac{(2 + S_j(A) + S_j(B))  \left( (1 + S_j(A))(1 + S_j(B)) - (1 + S_j(A \cup B))(1 + S_j(A \cap B)) \right)}{(1 + S_j(A \cup B))(1 + S_j(A \cap B))(1 + S_j(A))(1 + S_j(B))} }\\
\end{align*}


Additionally, we observe that:

\begin{align*}
&(1 + S_j(A))(1 + S_j(B)) - (1 + S_j(A \cup B))(1 + S_j(A \cap B))\\ 
&= \left[1 + S_j(A \cap B) + S_j(A \setminus B)\right] \left[1 + S_j(A \cap B) + S_j(B \setminus A)\right] \\
&\quad - \left[1 + S_j(A \cap B) + S_j(A \setminus B) + S_j(B - A)\right](1 + S_j(A \cap B)) \notag \\
&= S_j(A \setminus B) \times S_j(B \setminus A). \notag
\end{align*}

Therefore, we can substitute the equation in and have:

\begin{align*}
&\frac{1}{1 + S_j(A \cup B)} + \frac{1}{1 + S_j(A \cap B)} - \frac{1}{1 + S_j(A)} - \frac{1}{1 + S_j(B)}\\
& = \frac{(2+S_j(A)+S_j(B))(S_j({A \setminus B}))(S_j({B \setminus A}))}{(1 + S_j({A \cup B}))(1 + S_j({A \cap B}))(1 + S_j(A))(1 + S_j(B))}.
\end{align*}

Therefore we proofed that:
\begin{align*}
    &\mathbb{Q}_j(A \cup B) + \mathbb{Q}_j(A \cap B) - \mathbb{Q}_j(A) - \mathbb{Q}_j(B)\\
    &=\quad \frac{(2+S_j(A)+S_j(B))(S_j({A \setminus B}))(S_j({B \setminus A}))}{(1 + S_j({A \cup B}))(1 + S_j({A \cap B}))(1 + S_j(A))(1 + S_j(B))}.
\end{align*}
\end{proof}

\section{Proof of Theorem4.2}

We present the proof of Theorem 4.2 here:
    \begin{proof}
     We aim to demonstrate that the objective  
    \[
    \mathbb{Q}'(X) = \sum_{j=1}^M \mathbb{Q}_j(X) = \sum_{j=1}^M \frac{C_D(\mathbf{q}_j) + 1+ \sum_{t_i\in X} t_i.\beta\times w_{ij}}{C_D(\mathbf{q}_j) + 1 }
    \]
    Due to the linear nature of the target $\mathbb{Q}'(X)$, we can move the summation over $t_i.\beta$ from the inner layer to the outer layer, rearrange $t_i.\beta$, and combine elements based on the sets $(A\setminus B)$, $(B\setminus A)$, and $(A\cap B)$. Then, we prove that $\mathbb{Q}'(X)$ 
    satisfies the modular property. 
    
We can rewrite \( \mathbb{Q}'(X) \) as:
\[
\mathbb{Q}'(X) = M + \sum_{j=1}^M \frac{\sum_{t_i\in X} t_i.\beta\times w_{ij}}{C_D(\mathbf{q}_j) + 1}.
\]
Let us define \( H_i = \sum_{j=1}^M \frac{w_{ij}}{C_D(\mathbf{q}_j) + 1} \). Consequently, we have
\[
\mathbb{Q}'(X) =  M+\sum_{i=1}^N t_i.\beta\times H_i.
\]
Considering the left-hand side, we obtain:
\[
\mathbb{Q}'(A \cup B) + \mathbb{Q}'(A \cap B) = \sum_{t_k \in A \cup B} t_k.\beta\times H_k + \sum_{t_k \in A \cap B} t_k.\beta\times H_k.
\]
This can be expanded as:
\[
\sum_{t_k \in A \setminus B} t_k.\beta\times H_k + 2 \sum_{t_k \in A \cap B} t_k.\beta\times H_k + \sum_{t_k \in B \setminus A} t_k.\beta\times H_k.
\]
On the other hand, the right-hand side is:
\[
\mathbb{Q}'(A) + \mathbb{Q}'(B) = \sum_{t_k \in A} t_k.\beta\times H_k + \sum_{t_k \in B} t_k.\beta\times H_k.
\]
This also expands to:
\[
\sum_{t_k \in A \setminus B} t_k.\beta\times H_k + 2 \sum_{t_k \in A \cap B} t_k.\beta\times H_k + \sum_{t_k \in B \setminus A} t_k.\beta\times H_k.
\]
Therefore, we have
\[
\mathbb{Q}'(A) + \mathbb{Q}'(B) = \mathbb{Q}'(A \cup B) + \mathbb{Q}'(A \cap B).
\]
This equality demonstrates the modular property.
\end{proof}

\section{Proof of Corollary 1}
Here, we will prove our Corollary 1. When the attacker specifies which queries to insert and which to delete, our optimization goal becomes

\begin{align}
    \text{Max}: & : \sum_{\mathbf{q}_j \in \text{Insert}} \frac{C_D(\mathbf{q}_j) + 1 + \sum_{i=1}^N t_i \cdot \beta \times w_{ij}}{C_D(\mathbf{q}_j) + 1} \nonumber \\
    & \quad + \sum_{\mathbf{q}_j \in \text{Delete}} \frac{C_D(\mathbf{q}_j) + 1}{C_D(\mathbf{q}_j) + 1 + \sum_{i=1}^N t_i \cdot \beta \times w_{ij}} \nonumber \\
    & \text{s.t.} \quad \sum_{i=1}^N |t_i \cdot \beta| \leq K, \quad t_i \cdot \beta \in \{-1, 0, 1, \dots, K\} \notag
\end{align}

According to Theorem~\ref{The.51} and Theorem~\ref{The.03}, the first term of the optimization objective

\begin{equation}
\sum_{\mathbf{q}_j \in \text{Insert}} \frac{C_D(\mathbf{q}_j) + 1 + \sum_{i=1}^N t_i \cdot \beta \, w_{ij}}{C_D(\mathbf{q}_j) + 1},
\end{equation}

exhibits modular properties. In contrast, the second term of the optimization objective

\begin{equation}
\sum_{\mathbf{q}_j \in \text{Delete}} \frac{C_D(\mathbf{q}_j) + 1}{C_D(\mathbf{q}_j) + 1 + \sum_{i=1}^N t_i \cdot \beta \, w_{ij}},
\end{equation}

demonstrates supermodular characteristics. The combination exhibits supermodular characteristics. This aligns with the conditions outlined in~\cite{GreedIsGood}, where the objective is defined as the sum of a supermodular function and a submodular function with zero curvature. Furthermore, Bai et al.~\cite{GreedIsGood} have demonstrated that this linear combination does not impact the approximation ratio of the greedy algorithm when applied to such linear summation functions. As a result, the approximation ratio remains:

\begin{equation}
\mathbb{Q}(\Delta D) \geq (1 - \kappa) \mathbb{Q}(O).
\end{equation}

\section{Proof of Corollary 2}


We aim to demonstrate the following:

{1. Under the given conditions, the optimal solution does not include insertions.}

Consider the scenario where insertions are contemplated. The upper bound of the potential gain from insertions arises from the repeated insertion across $M$ queries, specifically denoted as $K \times M$. This gain is inferior compared to the cost incurred by deleting one or two tuples corresponding to a query $q_x$, represented as $C_D(q_j)$. Mathematically, this relationship can be expressed as:

\[
K \times M < C_D(q_j)
\]

Consequently, the optimal solution favors deletion over insertion, implying that the optimal solution does not incorporate insertions.

{2. Under the given conditions, Algorithm 1's will not choose to insert}

Algorithm 1 evaluates the remaining undeleted tuples at each step. Consider the $L$-th step of the algorithm, where we examine an undeleted tuple $t_i$. We use $D_L$ to represent the current database state and let $S_1$ denote the set of queries $q_j$ involving $t_i$ with $count(w_{i,j}\neq 0 )=1$, $S_2$ denote the set of queries $q_j$ involving $t_i$ with $count(w_{i,j}\neq 0 )=2$. The potential gain from repeatedly inserting tuples associated with queries in $S$ is given by:

\[
gainI = \sum_{q_j \in S_1} \frac{2 \cdot C_{D_L}(q_j) + 1}{C_{D_{L}}(q_j) + 1} + \sum_{q_j \in S_2} \frac{ C_{D_L}(q_j) + 1 + w_{i,j}}{C_{D_L}(q_j) + 1}
\]

On the other hand, if tuple $t_i$ is deleted, the corresponding gain is:

\[
gainD = \sum_{q_j \in S_1} \frac{C_{D_L}(q_j) + 1}{1} +  \sum_{q_j \in S_2} \frac{C_{D_L}(q_j) + 1}{C_{D_L}(q_j) +1 - w_{i,j}}
\]

Note that:

${\small \frac{C_{D_L}(q_j) + 1}{C_{D_L}(q_j) +1 - w_{i,j}} - \frac{ C_{D_L}(q_j) + 1 + w_{i,j}}{C_{D_L}(q_j) + 1} =  \frac{w_{i,j}}{C_{D_L}(q_j) +1 - w_{i,j}} - \frac{ w_{i,j}}{C_{D_L}(q_j) + 1} > 0 }$

and:


\[
\frac{C_{D_L}(q_j) + 1}{1} - \frac{2 \cdot C_{D_L}(q_j) + 1}{C_{D_L}(q_j) + 1} = C_{D_L}(q_j) - \frac{C_{D_L}(q_j)}{C_{D_L}(q_j) + 1} \geq 0
\]


We can compute the difference between the inner summations of the two groups separately, yielding:

\begin{align*}
& \sum_{q_j \in S_1} \frac{C_{D_L}(q_j) + 1}{1} 
 - \sum_{q_j \in S_1} \frac{2 \cdot C_{D_L}(q_j) + 1}{C_{D_L}(q_j) + 1}  \\
& = \sum_{q_j \in S_1} \left( \frac{C_{D_L}(q_j) + 1}{1}  
 - \frac{2 \cdot C_{D_L}(q_j) + 1}{C_{D_L}(q_j) + 1}   \right) > 0
\end{align*}

Additionally, for the second group, we have:


\begin{align*}
& \sum_{q_j \in S_2} \frac{C_{D_L}(q_j) + 1}{C_{D_L}(q_j) +1 - w_{i,j}}  
 - \sum_{q_j \in S_2} \frac{ C_{D_L}(q_j) + 1 + w_{i,j}}{C_{D_L}(q_j) + 1}  \\
& = \sum_{q_j \in S_2} \left( \frac{C_{D_L}(q_j) + 1}{C_{D_L}(q_j) +1 - w_{i,j}} 
 -  -\frac{ C_{D_L}(q_j) + 1 + w_{i,j}}{C_{D_L}(q_j) + 1}    \right) > 0
\end{align*}

Combining the results from both groups, we obtain:

\[
gainI < gainD
\]

This inequality indicates that the gain from insertion ($gainI$) is less than the gain from deletion ($gainD$). Therefore, Algorithm 1 will opt for deletion over insertion at each step. As a result, the algorithm inherently avoids insertions, effectively reducing the problem to an insert-only scenario under the given conditions.

\begin{equation}
\mathbb{Q}(\Delta D) \geq (1 - \kappa) \mathbb{Q}(O).
\end{equation}

\section{Countermeasure Study}
In this chapter, we will discuss how to defend against DACA attacks and present  additional potential defense mechanisms, analyzing their strengths and limitations.

\subsection{Noise Injection as a Defense Mechanism}

Given that worst-case attackers exploit all available information to degrade the performance of cardinality estimators towards oracle baselines, we propose a defense strategy that systematically perturbs estimator outputs to increase their statistical distance from database-specific oracles. Inspired by differential privacy mechanisms which use adequate noise to protect the key distribution~\cite{DP_dong2024continual,DP_dong2024instance,DP_sun2023confidence}, we investigate controlled noise injection to obscure data distribution characteristics while maintaining estimator utility. Formally, we modify estimator outputs as:

\begin{equation}
    \text{Est}_{\text{Noise}} = \text{Est} + |\alpha  \cdot \eta|
\end{equation}

\noindent where $\sigma$ denotes the standard deviation of estimator outputs, $\alpha$ controls noise intensity, and $\eta \sim \mathcal{N}(0, \sigma^2)$ is  Gaussian noise. This perturbation creates an $\alpha\sigma$-radius protection boundary around original estimates.

\begin{figure}[htbp]
    \centering
    \subfigure[IMDB-JOB Dataset \label{Fig:IMDB_NOISE}]{
        \includegraphics[width=0.4\columnwidth]{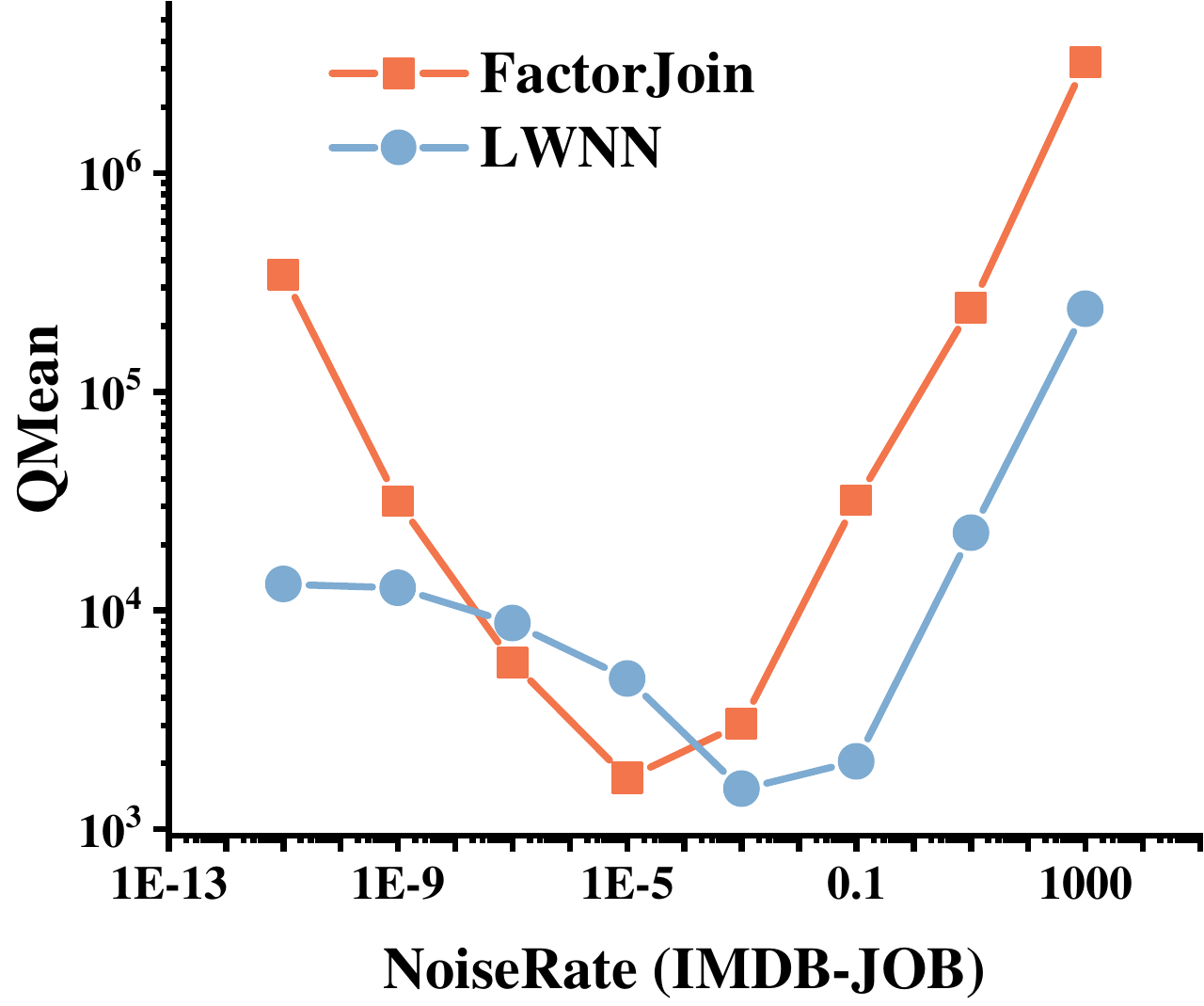}
    }
    \subfigure[STATS-CEB Dataset \label{Fig:STATS_NOISE}]{
        \includegraphics[width=0.4\columnwidth]{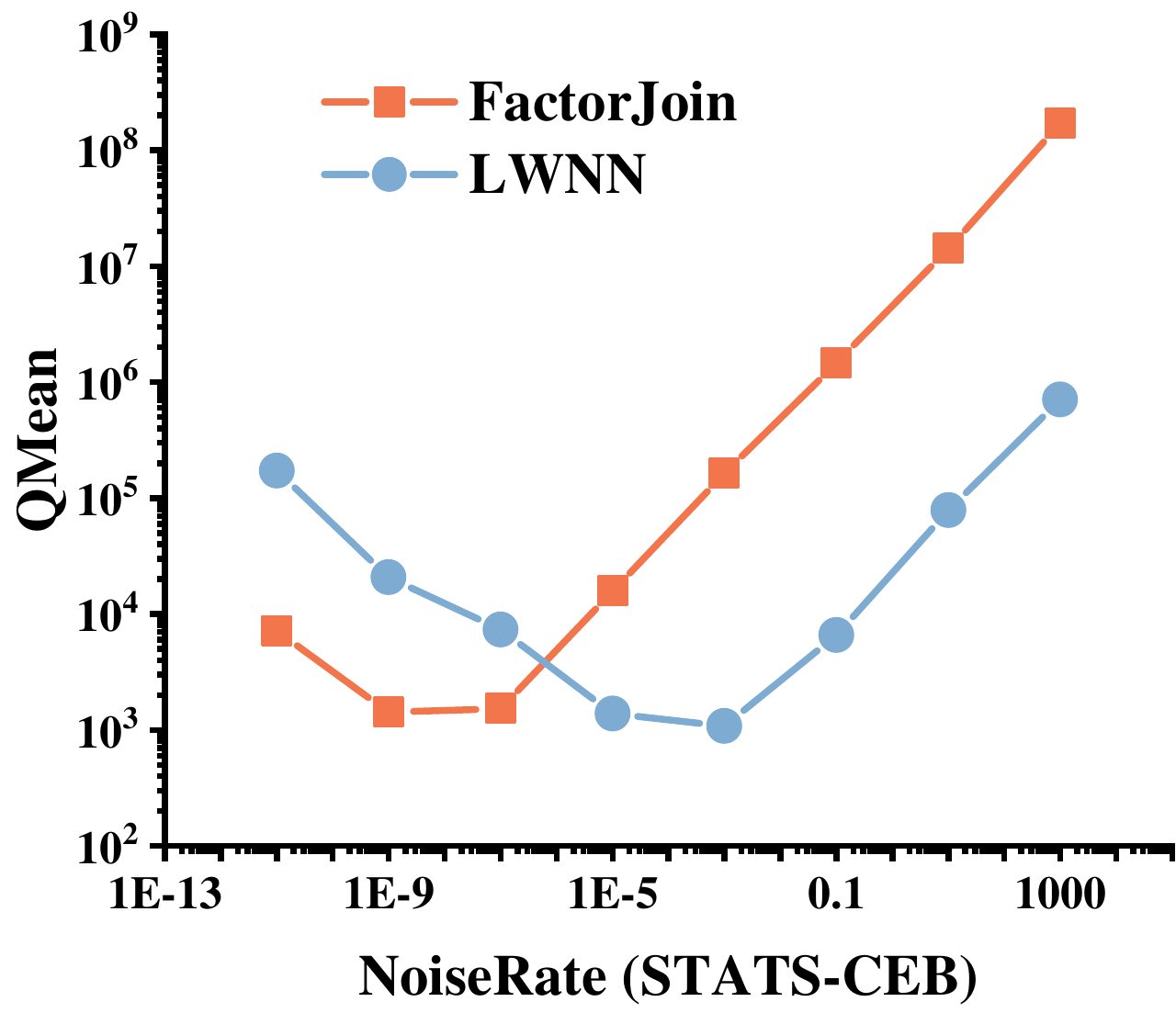}
    }
    \caption{Defense Performance via Noise Injection}
    \label{Exp.Noise}
    \vspace{-1.5em}
\end{figure}

Figure~\ref{Exp.Noise} demonstrates the defense effectiveness across different estimators and datasets. Our empirical analysis reveals three key findings: (1) Appropriate noise injection (QMean reduction up to 3 orders of magnitude) successfully mitigates worst-case performance degradation; (2) Optimal $\alpha$ values exhibit dataset- and estimator-dependent characteristics, complicating universal parameter selection; (3) Excessive noise ($\alpha > 1.0$) induces estimator collapse where defense-induced error exceeds attack impacts, highlighting the critical precision-robustness tradeoff.

\begin{yellowbox}
\noindent\textbf{Theoretical Implications:} Our findings suggest that adaptive noise calibration requires fundamental understanding of estimator-specific vulnerability profiles. Future work should establish: (1) Attack surface characterization through estimator sensitivity analysis; (2) Noise-response curves for automated parameter tuning; (3) Information-theoretic bounds for privacy-utility tradeoffs in cardinality estimation.
\end{yellowbox}

\end{document}